\documentclass[12pt]{article}
\usepackage[utf8]{inputenc}
\usepackage[a4paper,margin=1in]{geometry}
\usepackage{setspace}
\usepackage{algorithm}
\usepackage{algpseudocode}
\usepackage[dvipsnames,table,xcdraw]{xcolor}
\usepackage{graphicx}
\usepackage{xr}
\usepackage{hyperref}
\usepackage{subcaption}
\usepackage{mathtools}
\usepackage[toc]{appendix}
\usepackage{bm}
\usepackage{natbib}
\usepackage{etoolbox}
\usepackage{amsmath}
\usepackage{amssymb}
\usepackage{bm}
\usepackage{bbold}
\usepackage[normalem]{ulem}
\usepackage[labelfont=bf]{caption}

\newcommand{\bs}{\boldsymbol}

\makeatletter
\patchcmd{\@makecaption}
  {\parbox}
  {\advance\@tempdima-\fontdimen2} 
  {}{}
\makeatother  

\providecommand{\keywords}[1]
{
  \small	
  \textbf{Keywords ---} #1
}

\counterwithout{equation}{section}

\definecolor{dkgreen}{rgb}{0,0.6,0}
\definecolor{gray}{rgb}{0.5,0.5,0.5}
\definecolor{mauve}{rgb}{0.58,0,0.82}
\definecolor{black}{rgb}{0.0, 0.0, 0.0}
\definecolor{MyDarkBlue}{RGB}{51,127,153} 
\definecolor{MyLightBlue}{rgb}{0.2,0.2,1.0} 

\newcommand*\samethanks[1][\value{footnote}]{\footnotemark[#1]}

\title{Bayesian nonparametric clustering for spatio-temporal data, with an application to air pollution}

\author{
  Luca Aiello\thanks{Department of Economics, Università degli studi di Bergamo, Italy, \href{mailto:luca.aiello@unibg.it}{luca.aiello@unibg.it}},
  Raffaele Argiento\samethanks, 
  Sirio Legramanti\samethanks, 
  Lucia Paci\thanks{Department of Statistical Sciences, Università Cattolica del Sacro Cuore di Milano, Italy}
}

\date{}

\begin{document}

\maketitle

\begin{abstract}
    Air pollution is a major global health hazard, with fine particulate matter (PM10) linked to severe respiratory and cardiovascular diseases. Hence, analyzing and clustering spatio-temporal air quality data is crucial for understanding pollution dynamics and guiding policy interventions. This work provides a review of Bayesian nonparametric clustering methods, with a particular focus on their application to spatio-temporal data, which are ubiquitous in environmental sciences. We first introduce key modeling approaches for point-referenced spatio-temporal data, highlighting their flexibility in capturing complex spatial and temporal dependencies. We then review recent advancements in Bayesian clustering, focusing on spatial product partition models, which incorporate spatial structure into the clustering process. We illustrate the proposed methods on PM10 monitoring data from Northern Italy, demonstrating their ability to identify meaningful pollution patterns. This review highlights the potential of Bayesian nonparametric methods for environmental risk assessment and offers insights into future research directions in spatio-temporal clustering for public health and environmental science.
\end{abstract}


\section{Introduction}

Air pollution represents one of the most pressing environmental challenges of our time, threatening ecosystems and public health, both directly and indirectly. In fact -- besides directly harming the health of those who inhale polluted air -- air pollution can also be regarded as both a cause and a consequence of the wider climate change issue. In light of these growing risks, constant air quality monitoring and rigorous statistical analysis of the resulting data are crucial for identifying the most vulnerable areas and informing effective mitigation strategies. This requires statistical models that can manage uncertainty, capture complex relationships, and integrate diverse data.

Air quality measurements are inherently spatio-temporal, being recorded across both spatial and temporal dimensions. Starting from the first dimension, spatial data are conventionally classified into three types: point-referenced (geostatistical) data, which consist of measurements taken at specific locations (e.g., air quality levels at different monitoring stations); areal (lattice) data, where measurements are aggregated over defined spatial regions (e.g., land use rates by administrative areas); and point-pattern data, which refer to the spatial occurrence of events (e.g., the locations of earthquakes or wildlife sightings). For a comprehensive overview of spatial data, refer to \cite{banerjee2014hierarchical} and \cite{cressie2015statistics}. A similar distinction can be made for the temporal dimension, where time can be treated either as continuous (over $\mathbb{R}^{+}$ or a subinterval) or discrete (e.g., hourly, daily, etc.). In the latter case, it is important to clarify whether each measurement represents a count or an average over a time interval -- analogous to areal measurements in spatial data -- or a single point-in-time observation.

In the last three decades, the increasing availability of datasets indexed by both space and time has driven the development of stochastic models that take into account both the spatial and temporal structure, and possible interactions of the two. For instance, in air pollution studies, we are not only concerned with the spatial distribution of a pollutant but also with how this distribution changes over time. Bayesian spatio-temporal models play a crucial role in this scenario, providing a coherent probabilistic framework for quantifying uncertainty and incorporating both spatial and temporal dependencies. Bayesian models are also particularly effective for cluster analysis, which is a well-known challenge in modeling spatio-temporal data and, more generally, environmental data. Applications range from early studies identifying minefields or seismic faults \citep{dasgupta1998detecting} to more recent works on air quality monitoring \citep{cheam2017model}, food-searching behavior of sandhoppers \citep{ranalli2020model}, coastal rainfall patterns \citep{paton2020detecting}, and fish biodiversity \citep{vanhatalo2021spatiotemporal}, just to name a few. In many applications, the focus is on the cluster allocation of the observations, as well as on the patterns within each cluster.

More generally, cluster analysis refers to inferring a partition of the sample into subsets named \textit{clusters}, so that statistical units belonging to the same cluster are as similar as possible to each other, while being as different as possible from units in the other clusters. In other words, clustering aims at cohesion within clusters and separation across clusters. In the basic version of clustering, the abovementioned partition is a partition in the mathematical sense, i.e., a collection of non-overlapping subsets whose union reconstructs the whole sample. That is, each unit is assigned to one and only one cluster. However, more sophisticated versions of clustering allow, e.g., that the same unit belongs to more than one cluster \citep[overlapping clusters, ][]{palla2005uncovering}, and/or assign to each unit a degree \citep[fuzzy clustering, ][]{bezdek1984fcm} or probability of membership (Bayesian clustering, see Section~\ref{sec:bayesian_clustering}).

Clustering methods can be broadly classified into heuristic (algorithmic) and model-based (probabilistic) approaches. A well-known heuristic method is the \textit{k-means} algorithm, which partitions the observations into $k$ clusters by iteratively assigning each data point to the cluster with the closest mean. For a comprehensive discussion on \textit{k-means} and other distance-based clustering techniques, see \citet{sarang2023centroid}. Heuristic strategies have the advantage of being fast and capable of handling large volumes of data or high-dimensional datasets. On the other hand, they only return a point estimate of the abovementioned partition, without quantifying the uncertainty about the cluster membership of each unit. In contrast, model-based approaches \citep{fraley2002model} allow quantifying the uncertainty on the inferred partition by including a clustering structure in the statistical model. This also allows for addressing questions such as how many clusters are required, how to detect and treat outliers, and how covariates can inform the clustering. 

A natural approach to probabilistic clustering is based on mixture models, where observations are modeled by a convex combination (a linear combination with weights summing up to one) of densities, referred to as \textit{components} of the mixture. Such densities typically belong to the same parametric family, referred to as \textit{mixture kernel}. For more details on mixture models, see Section~\ref{sec:clustering}. Inference on the component weights and component-specific density parameters can then be carried out, based on the observed data, via either frequentist or Bayesian approaches. See \citet{bouveyron2019model} for a comprehensive overview of model-based clustering, and \citet{wade2023bayesian} and \citet{grazian2023review} for in-depth treatments of Bayesian approaches to it.

A key problem in mixture models is represented by specifying the number of components. This mirrors the problem of determining the number of clusters (we will see more about the relationship between clusters and components below). Note that, while no more than $n$ groups are needed for a finite sample of size $n$, at the population level we could, in principle, allow for infinitely many clusters. This makes the statistical clustering problem inherently nonparametric. Similarly, in mixture models, we can allow for an infinite or a finite, yet random, number of components, thus adopting a nonparametric approach. While infinite mixture models have long been investigated in Bayesian nonparametrics, only recently the case of a random number of components has been formally framed within the nonparametric approach \citep{argiento2022infinity}. In both cases, the mixture induces a prior on the partition such that --  in the large-sample limit -- the number of clusters is not bounded from above. The most notable examples of such priors include the so-called \textit{mixture of finite mixtures} (see, e.g., \cite{miller2018mixture}), which assumes a finite but random number of components, and the Dirichlet process (DP) mixture model \citep{ferguson1973bayesian,lo1984class}, which allows for an infinite number of components. Many other flexible generalizations have been proposed in the literature; for a comprehensive overview, we refer the reader to \citet{muller2015bayesian} and \citet{fruhwirth2019handbook}. 

When it comes to clustering spatial point-referenced data, a prominent Bayesian nonparametric method is the spatial Dirichlet Process \citep{gelfand2005bayesian}, arising as a probability-weighted collection of random spatial surfaces. This model was generalized by \citet{duan2007generalized} by incorporating spatially dependent mixture weights. Both of these spatial Dirichlet processes require multiple observations at each spatial location. As an alternative, the spatial stick-breaking process \citep{reich2007multivariate,rodriguez2011nonparametric} allows spatially-varying allocation probabilities without requiring the presence of replications. Additionally, the Dirichlet labeling process \citep{nguyen2011dirichlet} has been proposed as a flexible model for spatial clustering. Although these methods are highly appealing, their performance can be limited when covariates are available to aid in cluster identification. For clustering areal data, the seminal work of \cite{fernandez2002modelling} introduced a finite mixture model with a random number of components and spatially dependent mixing weights, which are modeled through Markov random fields. As for time series clustering, \citet{nieto2014bayesian} introduced a Bayesian nonparametric mixture model within the state-space framework, while \citet{fruhwirth2008model}  proposed a finite mixture of dynamic regression models.

Building on this literature, this paper demonstrates the potential of Bayesian nonparametric models for clustering spatio-temporal data, which are prevalent in environmental science, including but not limited to air quality monitoring. In particular, such methods allow us to quantify the uncertainty about the estimated partition and the cluster-specific parameters, to avoid pre-specifying the number of clusters, and to incorporate geographical coordinates, which are also typically available in environmental datasets. In particular, incorporating covariates yields more parsimonious and spatially cohesive -- and hence more interpretable -- clusters. 

The rest of the paper is organized as follows.  In Sections \ref{sec:spacetime} and \ref{sec:clustering}
we provide an overview of mixture modeling for cluster analysis, focusing on spatio-temporal data and Bayesian nonparametric methods. 
In particular, in Section~\ref{sec:sPPM}, we provide a detailed discussion of Product Partition Models \citep[PPM, ][]{hartigan1990partition} and their extensions to the spatio-temporal setting. We then illustrate this methodology on real-world air quality data (Section~\ref{sec:data_analysis}), with the aim of highlighting the wide applicability of such techniques for all applications involving spatio-temporal data. Finally, Section~\ref{sec:discussion} offers concluding remarks and avenues for future research, stressing the potential of Bayesian nonparametric methods for environmental data science.

\section{Models for spatio-temporal data}
\label{sec:spacetime}

Denote by $D \subseteq \mathbb{R}^2 $ a continuous domain, and let $\bs{s} \in D$ be a point within it. We then denote by $\left\{ Y_{t}(\bs{s}): {\bs{s} \in D}; \ {t=1,\dots,T} \right\}$ a spatio-temporal process that can be seen as a process generating a time series indexed by $t=1,\dots,T$ at each location $\bs{s} \in D$. For alternative definitions of spatio-temporal processes, see \citet{banerjee2014hierarchical}. Measurements at each location and time can be continuous, binary, or count data -- for example, pollution levels, species presence/absence, or their abundance, respectively.

A customary model for continuous spatio-temporal data is  
\begin{equation}\label{eq:gen_form}
    y_{t}(\bs{s}) = \bs{x}_{t}(\bs{s})^\top \bs{\beta}_{t}(\bs{s}) + w_{t}(\bs{s}) + \epsilon_{t}(\bs{s}),
\end{equation}
where $\bs{x}_{t}(\bs{s})^\top \bs{\beta}_{t}(\bs{s})$ represents the mean structure (i.e., the linear predictor) of $y_{t}(\bs{s})$ given the spatio-temporal covariates $\bs{x}_{t}(\bs{s})$, while $w_{t}(\bs{s})$ is a mean-zero process capturing the spatio-temporal dependence that is not explained by the covariates, and $\epsilon_{t}(\bs{s})$ is an error term -- typically a Gaussian white noise process -- accounting for the residual spatio-temporal variability in the data.

In principle, the regression coefficient $\bs{\beta}_{t}(\bs{s})$ may depend on both space and time. However, it can be simplified to $\bs{\beta}_{t}(\bs{s}) = \bs{\beta}$ \citep{rushworth2017adaptive,hefley2017dynamic,laurini2019spatio,wang2024spatiotemporal}, or at least to $\bs{\beta}_{t}(\bs{s}) = \bs{\beta}_{t}$ \citep{kottas2008modeling,berrocal2016identifying,lee2021clustered} or $\bs{\beta}_{t}(\bs{s}) = \bs{\beta}(\bs{s})$ \citep{torabi2014spatiotemporal,li2016comparison,peluso2020bayesian}. More details on the representation of $\bs{\beta}_{t}(\bs{s})$ can be found in \cite{banerjee2014hierarchical}. 

As for $w_{t}(\bs{s})$, different modeling choices are available in the literature, including the additive form $w_{t}(\bs{s}) = \lambda_{t} + \psi(\bs{s})$ \citep{waller1997hierarchical,knorr1998modelling,cameletti2011comparing}, where there is no shared information between space and time, and the multiplicative form $w_{t}(\bs{s}) = \lambda_{t} \psi(\bs{s})$ \citep{paci2013spatio}. In the last two cases, a popular approach for modeling $\psi(\bs{s})$ is via a Gaussian process \citep{williams2006gaussian} equipped with a spatial covariance structure. More recently, \citet{laurini2019spatio} and \citet{wan2021spatio} proposed to treat $w_{t}(\bs{s})$ as a pure spatial random effect, assuming $\lambda_t=\lambda$, constant over time. Another popular choice for $w_{t}(\bs{s})$, is to assume a dynamic evolution of the spatial process over time, namely
\begin{equation}\label{eq:din_rand_eff}
    w_{t}(\bs{s}) = \phi w_{t-1}(\bs{s}) + \nu_{t}(\bs{s}),
\end{equation}
where the spatial processes $\nu_{t}(\bs{s})$ are independent across times. Altogether, \eqref{eq:gen_form} and \eqref{eq:din_rand_eff} constitute the well-known dynamic linear model \citep{harrison1976bayesian,west2006bayesian}, often referred to as \textit{state-space model} in the time series literature. This approach allows for modeling temporal components such as trends, seasonal effects, and autoregressions. For example, \cite{stroud2001dynamic} applied this model to tropical rainfall levels and sea surface temperatures in the Atlantic Ocean. 

When clustering spatio-temporal data, the focus is typically on the spatio-temporal component $w_{t}(\bs{s})$.  This component can be seen as either $n$ time series $\bs{w}(\bs{s}_i)=(w_{1}(\bs{s}_i),\dots,w_{T}(\bs{s}_i))^\top$ for $i=1,\dots,n$, one for each location, or as $T$ spatial surfaces $\bs{w}_t = (w_{t}(\bs{s}_1),\dots,w_{t}(\bs{s}_n))^\top$, one for each time point. A first option would be to cluster together locations $\bs{s}_i$ and $\bs{s}_j$ if the corresponding latent time series take the same values over the whole time interval, i.e., $w_{t}(\bs{s}_i)=w_{t}(\bs{s}_j)$ for each $t=1,\ldots,T$; see, e.g., \citet{berrocal2016identifying} and \citet{nieto2014bayesian}. Note that we are not requiring that the observed time series $y_{t}(\bs{s}_i)$ and $y_{t}(\bs{s}_j)$ coincide along their whole trajectory -- which would be not realistic -- but only that the corresponding latent ones do. Alternatively, one may want to assign two time points $t$ and $t^\prime$ to the same cluster if the corresponding latent surfaces coincide at all the considered locations, i.e., $w_{t}(\bs{s}_i)=w_{t'}(\bs{s}_i)$ for each $i=1,\ldots,n$; see, e.g., \citet{gelfand2005bayesian} and \citet{page2022dependent}. A third option consists of clustering over time and space simultaneously, meaning that the pairs $(\bs{s}_i,t)$ and $(\bs{s}_j,t')$ are assigned to the same cluster if $w_{t}(\bs{s}_i) = w_{t^\prime}(\bs{s}_{j})$; see, e.g., \citet{duan2007generalized} and \citet{nguyen2011dirichlet}.

Finally, clustering can be based on the parameters governing the spatio-temporal random effect $w_{t}(\bs{s})$; see, e.g., \citealt{mastrantonio2019new}, \citealt{bucci2022clustering}, and \citealt{musau2022clustering}. For instance, if we assume that $w_{t}(\bs{s}_i) = \phi_i w_{t-1}(\bs{s}_i) + \nu_{t}(\bs{s}_i)$ with $\nu_t(\bs{s}_i) \sim \mathcal{N}(0,\tau_i^2)$ as in~\eqref{eq:din_rand_eff}, we can cluster the locations $\bs{s}_i$ based on the parameters $\bs{\theta}_i = (\phi_i, \tau_i^2)$. In the present work, we will follow this latter approach.

\section{Mixture models for cluster analysis} \label{sec:clustering}

In mixture modeling, each observation is assumed to come from one of the $ M\leq \infty$ mixture components. Namely, the model is given by 
\begin{equation} \label{eq:eq_mixture}
    f(y \mid P) = \int_{\Theta} f(y \mid \theta) P(d\theta) = \sum_{m = 1}^M \pi_{m} f(y \mid \gamma_{m}),
\end{equation}
where $\{f(y \mid \theta): \theta \in \Theta\}$ is a parametric family of densities on $\mathcal{Y}$, referred to as \emph{mixture kernel}, while $P(\cdot) = \sum_{m = 1}^M \pi_{m} \delta_{\bs{\gamma}_m}(\cdot)$ is an almost surely discrete measure on $\Theta$, supported on $\{\gamma_m \in \Theta: m=1,\dots,M\}$, and is referred to as the \emph{mixing measure}. A Bayesian mixture model is completed by assigning a prior distribution on $P$. However, instead of resorting to the stochastic process theory, we can observe that the discrete measure $P$ is completely characterized by the triplet $(M, \bs{\pi}, \bs{\gamma})$, where $\bs{\pi} = (\pi_1, \dots, \pi_M)$ and $\bs{\gamma} = (\gamma_1, \dots, \gamma_M)$. So, a prior on $P$ can be assigned by first choosing a probability mass function for $M$ (or alternatively setting it to either a finite value or to infinity; see below), and then specifying a joint law for $\bs{\pi}$ and $\bm{\gamma}$. 

Regarding the number of components $M$, three approaches are prevalent. A first option requires fixing the number of components $M$ to a finite value and specifying a symmetric Dirichlet distribution for the weights, i.e., $\bs{\pi}\mid M \sim \mbox{Dirichlet}_M(\tilde{\alpha}, \dots, \tilde{\alpha})$, where the hyperparameter $\tilde{\alpha}$ regulates the prior information about the relative sizes of the mixing weights, e.g., small values of $\tilde\alpha$ favor lower entropy $\pi$'s. An alternative is to treat $M$ as a random variable with prior $M \sim q_M$, while still assuming $\bs{\pi}\mid M \sim \mbox{Dirichlet}_M(\tilde{\alpha}, \dots, \tilde{\alpha})$. This defines a mixture model with a random number of components, which is referred to as a \emph{mixture of finite mixtures} \citep{miller2018mixture,fruhwirth2021generalized}. In this setting, popular choices for $q_M$ are a Poisson distribution \citep{stephens2000bayesian,nobile2004posterior}, a uniform distribution over $\{1, \dots, M_{\max}\}$ \citep{miller2018mixture}, or a Negative Binomial \citep{grazian2020loss}. Finally, one can set the number of components to infinity ($ M = \infty $). In this case, the arguably most popular prior for $\bs{\pi}$ is the stick-breaking construction \citep{sethuraman1994constructive}, also known as the $ \text{GEM}(\alpha) $ distribution \citep{griffiths1974properties}. In particular, the weights are defined as $ \pi_1 = \tilde{\pi}_1 $, $ \pi_m = \tilde{\pi}_m \prod_{j=1}^{m-1} (1 - \tilde{\pi}_j) $ for $ m \geq 2 $, with $ \tilde{\pi}_m \overset{\text{iid}}{\sim} \text{Beta}(1, \alpha) $. Under this formulation, the random probability measure $ P $ is a Dirichlet Process \citep[DP, ][]{ferguson1973bayesian}, and model \eqref{eq:mod_clust} corresponds to the well-known DP mixture model \citep{lo1984class}. Note that, if $\tilde{\alpha}=\alpha/M$, then, when $M \to \infty$ the finite mixture models described above converge in law to a DP($\alpha$) mixture model. This asymptotic property has often been exploited to facilitate posterior computation under DP mixture models. Moreover, working with a large value of $M$ endows finite mixture models with appealing theoretical and practical properties \citep[see, e.g., the \emph{sparse mixture} modeling in ][]{rousseau2011asymptotic}. We also mention that, when using the DP, a hyperprior is often placed on $\alpha$ \citep[See, e.g.][]{escobar1995bayesian} in order to make the posterior inference on the number of clusters and the clustering itself more robust.

Finally, a prior distribution on the component-specific parameters must be assigned. Usually, given $M$, the parameters $\gamma_{m} \in \Theta \subset \mathbb{R}^d$, with $m = 1, \dots, M$, are assumed to be independent and identically distributed according to a non-atomic prior $P_0$. For computational convenience, a common choice for $P_0$ is a conjugate prior to the kernel $f(\bs{y} \mid \gamma)$. The hyperparameters of the base measure $P_0$ can be specified based on prior knowledge, set empirically, or inferred through additional hyperpriors. Alternatively, data-dependent or non-informative priors can be used for $P_0$ \citep{rousseau2019bayesian}. Finally, to encourage well-separated components, the independence assumption on the atoms $\gamma_m$ can be relaxed by employing repulsive priors \citep{petralia2012repulsive,xie2020bayesian,beraha2025}, determinantal point processes \citep{xu2016bayesian}, or non-local priors \citep{fuquene2019choosing}.

The choice of the mixture kernel is also crucial -- although often overlooked -- and depends on the nature of the data, as well as on the objectives of the analysis. For continuous data, while Gaussian kernels stand out as the most popular choice, skewness or outlier robustness can be achieved by using, e.g., multivariate skew-Normal or skew-t distributions \citep{fruhwirth2010bayesian,lee2014finite}, shifted asymmetric Laplace distributions \citep{franczak2013mixtures}, or Normal inverse Gaussian distributions \citep{o2016clustering}. For categorical data, one may employ latent class models, which rely on mixtures of Bernoulli or multinomial distributions \citep{goodman1974exploratory,argiento2024}. Finally, for count data, mixtures of Poisson \citep{karlis2005mixed,krnjajic2008parametric}, negative binomial \citep{liu2024shared}, or zero-inflated Poisson and negative-binomial distributions are useful for handling sparse counts \citep{wu2022nonparametric}. 

\subsection{Bayesian clustering}
\label{sec:bayesian_clustering}

As is customary in mixture modeling, a hierarchical representation of the mixing measure can be adopted to facilitate computations. Specifically, we introduce an allocation vector $\bs{c} = (c_1,\dots,c_n)^\top$, where each $c_i \in \{1,\ldots,K\}$ and $c_i=k$ means that the $i$-th unit is assigned to the $k$-th cluster. Such a vector induces a partition of the sample (or, equivalently, of the first $n$ integers) into $K$ clusters. We denote such a partition with $\rho_n \vcentcolon= \{S_1, \dots, S_K\}$, where $S_k=\{i:c_i=k\}$ represents the $k$-th cluster. While each allocation vector induces a unique partition, the converse does not hold. In fact, the same partition can be induced by different allocation vectors. As a simple example, let $n$ be even and consider the case where the first half of the units are assigned to cluster~1 and the second half to cluster~2. This obviously induces the same partition as the allocation assigning the first half of the sample to cluster~2, and the second half to cluster~1. This phenomenon -- known as \emph{label switching} \citep{stephens2000dealing} -- must be accounted for in many clustering methods. Usually, the allocation variables $c_1,\ldots,c_n$ are assumed to be conditionally independent given the mixture weights $\bs{\pi}$, with $\pi_m = \Pr(c_i = m \mid \bs{\pi})$ for all $m$. In particular, each allocation variable follows a categorical distribution, namely $c_i \mid \bs{\pi} \overset{\text{iid}}{\sim} \text{Multinomial}_M(1; \bs{\pi})$. Framing the sampling model in equation~\eqref{eq:eq_mixture} within the Bayesian setting involves placing a prior on $P$, that is, a prior on an almost surely discrete probability measure. This task generally requires advanced probabilistic tools, but -- thanks to the introduction of the allocation vector $\bs{c}$ -- the model specification can be conveniently simplified as follows:
\begin{equation}\label{eq:mod_clust}
    \begin{aligned}
        y_i \mid c_i, \bs{\gamma} &\stackrel{\text{ind}}{\sim} f(y \mid \bs{\gamma}_{c_i}), &\quad &i = 1, \dots, n, \\
        c_i \mid \bs{\pi} &\overset{\text{iid}}{\sim} \text{Multinomial}_M(1; \bs{\pi}), &\quad &i = 1, \dots, n, \\
        \gamma_m \mid M &\overset{\text{iid}}{\sim} P_0, &\quad &m = 1,\ldots, M,\\
        \bs{\pi} \mid M &\sim p(\bs{\pi}),\\
        M &\sim q_M.
    \end{aligned}
\end{equation}
The model formulation in \eqref{eq:mod_clust} can be viewed as a generative model by reading it from bottom to top. Specifically, after sampling or fixing the number of components $M$, we can sample the mixture weights $\bs{\pi}$ and the component-specific parameters $\gamma_m$ $(m=1,\ldots,M)$, assign each unit $i$ to a component $c_i$, and finally sample the observation $y_i$ from the corresponding component of the mixture. Note that the last three lines of \eqref{eq:mod_clust} correspond to specifying a prior on the mixing measure $P$.

Naturally, when sampling $n < \infty$ observations, not all of the $M$ components will necessarily be sampled from. This is evident when $M = \infty$, but it is also possible even when $M$ is finite, especially if $M$ is large. In other words, $M$ denotes the total number of components at a population level, i.e.,  the maximum number of possible clusters as $n \to \infty$. In contrast, $K \leq M$ refers to the number of non-empty clusters, i.e., those clusters that contain at least one of the $n < \infty$ observations \citep[see, e.g.,][]{argiento2022infinity}. Obviously, $K$ is also upper bounded by $n$, while $M$ can be larger than $n$ and even infinite. \cite{nobile2004posterior} emphasizes the distinction between $K$ and $M$, observing that -- when $M$ is random -- the posterior distribution of $M$ might concentrate on higher values with respect to the posterior of~$K$. In contrast, when $M$ is fixed -- either to a finite value or to infinity -- the posterior inference focuses on $K$ and on the other parameters defining the mixing measure, $\bs{\pi}$ and $\bs{\gamma}$. 

A crucial point is that specifying a prior for the mixing measure $P$ in \eqref{eq:eq_mixture} -- e.g., through the last three lines of the hierarchical model in \eqref{eq:mod_clust} -- automatically induces a prior on the partition  $\rho_n$. To shed light on this, let us first consider the representation $\theta_i = \gamma_{c_i}$ for $i = 1, \dots, n$, so that $\theta_i \mid P \stackrel{\text{iid}}{\sim} P$. Since $P$ is almost surely discrete, ties can occur among the $\theta_i$'s. We denote by $\theta_1^\ast, \ldots, \theta_K^\ast$ the $K$ unique values among $\theta_1, \dots, \theta_n$. In model-based clustering, the parameters of interest are the clustering itself (i.e., $\rho_n$) and the cluster-specific parameters (i.e., $\theta_1^\ast, \ldots, \theta_K^\ast$). Given the data $\bs{y} = \{y_1, \ldots, y_n\}$, a direct Bayesian approach would require hierarchically specifying the joint distribution $p(\bs{y}, \rho_n, \theta_1^\ast, \dots, \theta_K^\ast) = p(\bs{y} \mid \rho_n, \theta_1^\ast, \dots, \theta_K^\ast)\, p(\theta_1^\ast, \dots, \theta_K^\ast \mid \rho_n)\, \pi(\rho_n)$. However, an indirect approach based on the hierarchical model in \eqref{eq:mod_clust} is usually adopted. This corresponds to defining the joint distribution $p(\bs{y}, \bs{c}, \bs{\pi}, \bs{\gamma}, M)$, where the allocation vector $\bs{c}$ induces the partition $\rho_n$, and the triplet $(\bs{\pi}, \bs{\gamma}, M)$ corresponds to specifies the mixing measure $P$. Thus, model~\eqref{eq:mod_clust} implicitly defines the joint distribution $p(\bs{y}, \rho_n, P)$. 

In order to obtain the clustering model, the state space is augmented by the cluster-specific parameters $(\theta_1^\ast, \dots, \theta_K^\ast)$ and then the mixing measure $P$  is integrated out from the joint distribution $p(\bs{y}, \rho_n, \theta_1^\ast, \dots, \theta_K^\ast, P)$.  It is worth mentioning that the prior on $\rho_n$, that results from marginalizing over $P$, is known as the \emph{Exchangeable Partition Probability Function} (EPPF). Summing up, after the augmentation and the marginalization, deferring all the mathematical details to Section~\ref{app:clust_mix} in the Appendices, the hierarchical mixture model in \eqref{eq:mod_clust} can be rewritten as:
\begin{align}
\label{eq:mod_clust_eppf}
    \bs{y} \mid \rho_n, \theta_1^\ast, \dots, \theta_K^\ast &\sim \prod_{k=1}^K \prod_{i \in S_k} f(y_i \mid \theta_k^\ast), \nonumber \\
    \theta_1^\ast, \dots, \theta_K^\ast \mid \rho_n &\stackrel{iid}{\sim} P_0, \\
    \rho_n &\sim \text{EPPF}(n_1, \dots, n_K). \nonumber
\end{align}
Note that the EPPF depends only on the size of the clusters $n_1, \dots, n_K$. Also, note that, given a suitable EPPF, there exists an almost surely discrete random probability measure $P$, known as a \emph{Species Sampling Model} \citep{pitman1996}, such that the model can be expressed as in \eqref{eq:mod_clust}.

Model~\eqref{eq:mod_clust_eppf} highlights some advantages of the model-based Bayesian approach to clustering. Unlike heuristic algorithms, which yield a single estimated partition $\hat{\rho}_n$, Bayesian methods treat the partition $\rho_n$ as the object of interest and enable formal inference based on its posterior distribution, $p(\rho_n \mid \bs{y})$. By modeling the observed data $\bs{y}$ conditional on $\rho_n$, and assigning a prior distribution over $\rho_n$ that reflects the analyst's prior beliefs on the clustering structure (e.g., on the number or size of clusters), Bayesian cluster analysis provides a principled way to quantify uncertainty about $\rho_n$ after observing the data.

There are several options for the EPPF in \eqref{eq:mod_clust_eppf}, i.e., for the prior on $ p(\rho_n) $; for a comprehensive review, see, e.g., \cite{quintana2006predictive}. In particular, when $M<\infty$ and we choose a $\mbox{Dirichlet}_M(\tilde{\alpha}, \dots, \tilde{\alpha})$ prior for the weights $\bs{\pi}$ and a $\text{Multinomial}_M(1; \bs{\pi})$ for the cluster labels $c_i$, 
\begin{equation}
\mbox{EPPF}(n_1, \dots, n_K) = V(n, K) \prod_{k=1}^K \frac{\Gamma(\tilde{\alpha} + n_k)}{\Gamma(\tilde{\alpha})},
    \label{eq:eppf_of_finite_dir}
\end{equation}
where the normalizing constant $V(n,K)$ depends on the prior on $M$; see \cite{argiento2022infinity} for more details. Instead, when $M=\infty$ and we opt for a DP($\alpha$) as prior for $P$, the EPPF is 
\begin{equation}
\label{eq:eppf_infinite_dp}
    \mbox{EPPF}(n_1, \dots, n_K) = \frac{\Gamma(\alpha)}{\Gamma(\alpha + n)} \alpha^K \prod_{k=1}^K \Gamma(n_k).
\end{equation}
In view of Section~\ref{sec:sPPM} -- where we will incorporate spatial covariates into the prior for partition $\rho_n$ -- it is crucial to observe that, in both cases above, the EPPF assumes the form of a Product Partition Model \citep[PPM, ][]{hartigan1990partition}, where the probability of each cluster $S_k$ is proportional to a function $C(S_k)$  -- referred as \emph{cohesion function} -- that measures how likely the elements of~$S_k$ are to be clustered together, a priori:
\begin{equation}\label{eq:PPM}
    p\left(\rho_n\right) \propto \prod_{k=1}^{K} C\left(S_k\right).
\end{equation} 
Trivially, the cohesion function of the finite Dirichlet process in \eqref{eq:eppf_of_finite_dir} is $C(S_k)=\Gamma(\tilde\alpha)/\Gamma(\tilde\alpha+n_k)$, while the EPPF induced by the Dirichlet process in \eqref{eq:eppf_infinite_dp} is $C(S_k)=\alpha \Gamma(n_k)$. Note that, in these two notable cases, the cohesion function only depends on cluster $S_k$ through its cardinality $n_k$. For details on the specific cohesion function rising under the Bayesian nonparametric framework, e.g., Gibbs-type priors, see \citet{gnedin2006exchangeable}, \citet{lijoi2008bayesian}, \citet{gnedin2010species}, and \citet{de2015gibbs}.

\subsection{Posterior inference}\label{subsec:post_inf}

Sampling from the posterior distribution of the cluster assignments and cluster-specific parameters using Markov chain Monte Carlo (MCMC) methods is relatively straightforward when the number of components is fixed and finite. In these cases, the algorithm often simplifies to a Gibbs sampler with conjugate full conditionals. This situation arises, for instance, when the base measure $P_0$ and the mixture kernel are conjugate and the weights are drawn from a Dirichlet distribution. 

In contrast, posterior sampling becomes more challenging when the number of components is either infinite or finite but random (as in mixtures of finite mixtures). In particular, the latter case requires a transdimensional algorithm that accommodates jumps between parameter spaces of different dimensions (according to the number of mixing components). An early breakthrough was made by \citet{green1995} and \citet{richardson1997bayesian}, who introduced the Reversible Jump MCMC algorithm for univariate Gaussian mixtures. This algorithm is quite popular, but it requires the design of good reversible jump moves, posing challenges in applications, particularly with high-dimensional parameter spaces.

On the other hand, posterior samplers for infinite mixture models can be classified into two groups: (i) \emph{conditional algorithms}, providing full Bayesian inference on both mixing parameters and the clustering structure \citep{papaspiliopoulos2008retrospective, kalli2011slice}; and (ii) \emph{marginal algorithms}, which simplify computation by integrating out the mixture parameters, focusing solely on clustering \citep{maceachern1998estimating}. Early progress was marked by the introduction of the Pólya urn Gibbs sampler \citep{escobar1994estimating, maceachern1994estimating, escobar1995bayesian, maceachern1998computational}, followed by the application of the stick-breaking representation of the DP by \cite{ishwaran2001gibbs}, which enabled posterior inference via Gibbs sampling. 

Recently, exploiting the link between finite and infinite mixture models, algorithms developed for Bayesian nonparametric models have been adapted to finite mixtures. Examples include the Chinese restaurant process sampler in \citet{miller2018mixture}, the telescoping sampling developed by \citet{fruhwirth2021generalized}, and the two augmented Gibbs samplers proposed by \citet{argiento2022infinity}. These advances have improved both the computational efficiency and the real-world applicability of finite mixture models. 

All the MCMC sampling schemes mentioned above return a sample from the posterior distribution over the entire space of partitions,  namely from $p(\rho_n,\theta_1^\ast,\dots, \theta_K^\ast|\bs{y})$, thus allowing to quantify the uncertainty of the clustering structure given the data. However, a natural issue in this setting is how to summarize such a posterior distribution, for example by providing an appropriate posterior point estimate of the clustering structure and of the cluster-specific parameters based on the MCMC output. The problem is twofold: first, in order to estimate the clustering and the group-specific parameters, the mixture model must be identified to avoid label switching \citep{MCMC_label_switch}; second, the high dimension of the partition space and the fact that many of these partitions are usually quite similar (e.g., differing only in the assignment of a few data points) make the posterior spread out across a large number of partitions. 

A possible solution to get a point estimate $\hat{\rho}_n$ of the partition from its posterior samples is to assign data points to the cluster with maximum posterior probability. More generally, from a decision-theoretic perspective, the goal is to find the partition minimizing the expected loss under the posterior. In this framework, the maximum a posteriori described above is optimal under a binary loss, but may not fully explore the partition space or account for partial similarities between partitions. An alternative approach is the Binder loss \citep{binder1978bayesian}, which imposes a penalty when pairs of observations that should be grouped together are placed in different clusters, and when pairs that should be separated are incorrectly clustered together. \citet{wade2018bayesian} instead proposes using the Variation of Information (VI) loss \citep{meilua2007comparing}, which compares the information shared between two partitions. Although the VI loss is more computationally demanding and is sensitive to the cluster initialization, it provides a more nuanced comparison. In practice, the posterior point estimate of the partition based on a loss function can be obtained using a greedy stochastic search approach \citep{dahl2022search}. This method minimizes a chosen loss function, such as the Binder loss or the VI, given a posterior Monte Carlo sample. This algorithm is implemented in the \texttt{R} package \texttt{salso} \citep{salso}. 

Another key aspect in Bayesian clustering is the estimation of the cluster-specific parameters. Typically, a posterior summary -- such as a posterior mean or median -- is first computed for each observation-specific parameter $\bs{\theta}_i$. These estimates are then grouped according to the estimated clustering and averaged within each cluster \citep{molitor2010bayesian}, thus obtaining a posterior summary of the cluster-specific parameters $\bs{\theta}^\ast_k$. Alternatively, two other approaches can be used to estimate cluster-specific parameters. The first one involves extracting the cluster-specific parameter values directly from the final posterior sample, assuming that the algorithm has reached convergence. This provides a snapshot of the clustering structure at a specific point, rather than an averaged summary over the posterior distribution. The second approach consists of estimating all the model parameters while conditioned on the estimated partition \citep{argiento2024}. This strategy ensures that the parameter estimation aligns with the most probable clustering structure, offering a characterization of cluster properties based on a single, representative partition.

\section{Spatial Product Partition Model} \label{sec:sPPM}

The PPM formulation of the prior distribution on the partition $\rho_n$ introduced in equation~\eqref{eq:PPM} can be extended to directly incorporate covariate information into the clustering process. This extension, known as PPM with covariates (PPMx), was first introduced by \citep{muller2011product} and -- when the covariates are spatial coordinates -- is also referred to as spatial PPM \citep[sPPM,][]{page2016spatial}. In general, PPMx is defined by augmenting the PPM with an additional factor for each cluster that induces the desired dependence on the covariates. Namely, let $s_k^{\ast}=\{\bs{s}_i: \ i \in S_k \}$ be the covariates (e.g., the spatial coordinates) for the units in cluster $k$, and denote with $g(\cdot)$ a non-negative function -- referred to as \emph{similarity function} -- taking higher values on sets of covariates judged to be more similar. The prior on the partition $\rho_n$ is then conditioned on the value taken by $\{\bs{s}_1,\dots,\bs{s}_n\}$. Specifically, the PPM in \eqref{eq:PPM} becomes 
\begin{equation}\label{eqn:spatialPPM}
     p\left( \rho_n \mid \bs{s}_1,\dots,\bs{s}_n \right) \propto \prod_{k=1}^{K} g(s_k^{\ast}) C\left(S_k\right).
\end{equation}
In PPMx and sPPM, the specification of the similarity function $g(\cdot)$ in \eqref{eqn:spatialPPM} plays a critical role in how data points are grouped into clusters. 

A first option for the similarity function, proposed by \cite{page2016spatial}, includes a term for singleton clusters, $\mathbb{1}\left\{n_k = 1\right\}$, and a term that depends on the sum $\mathcal{D}_{k}$ of the distances between the centroid of the $k$-th cluster and each point of the same cluster. Namely,
\begin{equation} \label{eq:similarity1}
     g_1\left(s_{k}^{\ast}\right) = \mathbb{1}\left\{n_k = 1\right\} + \frac{1}{\Gamma(\omega\mathcal{D}_k) \mathbb{1}\left\{\mathcal{D}_k \geq 1 \right\} + \mathcal{D}_k \mathbb{1}\left\{\mathcal{D}_k < 1 \right\}} \mathbb{1}\left\{n_k > 1\right\},
\end{equation}
where $\mathbb{1}\left\{\cdot\right\}$ denotes the indicator function. This formulation is intuitive in the context of spatial modeling, since the similarity $g_1(s^\ast_k)$ is large for small values of $\mathcal{D}_k$, i.e., when the points in the $k$-th cluster are close to each other. The scalar $\omega$ serves as a tuning parameter, regulating how much high within-cluster distances are penalized.

An alternative similarity function, also by \cite{page2016spatial}, is based on pairwise distances within a cluster, with a threshold parameter $a$. Specifically, 
\begin{equation} \label{eq:similarity2}
    g_2\left(s_{k}^{\ast}\right) = \prod_{i,j \in S_{k}} \mathbb{1} \left\{ d(\bs{s}_i, \bs{s}_{j}) \leq a \right\}.
\end{equation}
This approach sets a stricter condition for similarity, by considering all the $n_k(n_k-1)/2$ unordered pairs of units assigned to cluster $k$, rather than just the $n_k$ pairs considered by the $\mathcal{D}_k$ in \eqref{eq:similarity1}.

More probabilistic options are also available, e.g., in \cite{muller2011product}. These approaches employ a density function $q(\bs{s}_i \mid \bs{\xi}_k)$ and integrate out the cluster-specific parameters $\bs{\xi}_k$ with respect to either their prior predictive distribution $q\left(\bs{\xi}_{k} \right)$ or their posterior predictive one, $q\left(\bs{\xi}_{k} \mid s_{k}^{\ast} \right)$. In both cases, we obtain a function of $s_{k}^{\ast}$, that is, respectively,
\begin{align}
    g_3\left(s_{k}^{\ast}\right) &= \int \prod_{\substack{i \in S_{k} }} q\left(\bs{s}_{i} \mid \bs{\xi}_{k}\right) q\left(\bs{\xi}_{k}\right) d \bs{\xi}_{k},\label{eq:similarity3} \\ 
    g_4\left(s_{k}^{\ast}\right) &= \int \prod_{\substack{i \in S_{k} }} q\left(\bs{s}_{i} \mid \bs{\xi}_{k}\right) q\left(\bs{\xi}_{k} \mid s_{k}^{\ast} \right) d \bs{\xi}_{k}. \label{eq:similarity4}
\end{align}
This approach is especially valuable in spatial modeling, as it accounts not only for the distances between points, but also for the underlying spatial distribution patterns that may emerge. 

A popular choice for $q(\bs{s}_i \mid \bs{\xi}_k)$ in \eqref{eq:similarity3}-\eqref{eq:similarity4} is the multivariate Normal density with hyperparameters $\bs{\xi}_k =  (\bs{m}_k, V_k)$ representing the mean vector and the covariance matrix of the spatial locations; $ q\left(\bs{\xi}_{k}\right)$ is then typically set to the conjugate Normal-Inverse-Wishart distribution. This choice -- which is the one that we employ in our real-world data analysis in Section~\ref{sec:data_analysis} -- yields a closed form for \eqref{eq:similarity3}, which is detailed in the Appendices, Section~\ref{app:similarity}. 

In summary, the choice of the similarity function in the sPPM has a strong impact on the cluster formation and interpretation. By carefully specifying this function, we can ensure that clusters reflect meaningful spatial patterns. This is essential in fields like environmental sciences and public health, in which spatial relationships play a crucial role.

\subsection{Posterior sampling under the sPPM}
\label{subsec:PPMx_sampling}

Posterior sampling for the sPPM -- i.e., for model~\eqref{eq:mod_clust_eppf} with a prior on the partition $\rho_n$ that incorporates spatial information as specified in equation~\eqref{eqn:spatialPPM} -- can be carried out by adapting the algorithms for mixture models mentioned in Section~\ref{subsec:post_inf}, as described in \cite{muller2011product} and \cite{page2016spatial}.

Namely, at each iteration, our Gibbs sampler assigns each statistical unit $i$ to cluster $k$ with a probability that is informed by the likelihood of the data $y_i$, the cohesion between observations in cluster $k$ (with and without including $i$), and the prior on the partition defined by the PPM. Once the partition is updated, the next step involves updating the parameters associated with each cluster and the overall likelihood function. These parameters, which may include the mean, variance, and autoregressive coefficients for each cluster, are sampled from their full conditional distributions, i.e., their distribution conditional on all the other parameters, at their most recent update. The sampling is performed based on the data assigned to each cluster, reflecting the local information captured by the partition. 

As a result, our algorithm returns samples from the posterior distribution of the clustering configurations and cluster-specific parameters. This allows us to quantify the uncertainty in both the partition and the other model parameters. As mentioned in Section~\ref{subsec:post_inf}, a posterior point estimate of the partition can then be selected by minimizing a posterior loss function, such as the VI, or by choosing the clustering configuration that best fits the data according to the posterior samples. An example of this procedure -- tailored to the model employed in Section~\ref{sec:data_analysis} for analyzing air quality data -- is described in detail in Algorithm~\ref{alg:sppm_gs}.

\begin{algorithm}
    \caption{Gibbs Sampling for a sPPM model on the time-series parameters.}\label{alg:sppm_gs}
    \begin{algorithmic}[1]
    \footnotesize
        \State update, for $i=1,\dots,n$, $\bs{\beta}_i \mid \sigma_{i}^{2}, \bs{\theta}_i,\bs{\Sigma}_{\bs{\beta}}, \bs{Z}, \bs{y}_{i}$ from 
        \begin{equation*}
            \mathcal{N}_p\left(\bs{V}_{\bs{\beta}_{i}} \bs{Z}^\top \bs{Q}_{i}^{-1} \bs{y}_{i},\bs{V}_{\bs{\beta}_{i}}\right),
        \end{equation*}
        \Statex where $\bs{Q}_{i} = \sigma_{i}^{2} \bs{I} + \bs{R}(\bs{\theta}_i)$ and $\bs{V}_{\bs{\beta}_i} = \left( \bs{Z}^\top \bs{Q}_{i}^{-1} \bs{Z} + \bs{\Sigma}_{\bs{\beta}}^{-1}\right)^{-1}$
        \State update, for $i=1,\dots,n$, $\bs{w}_i \mid \sigma_{i}^{2}, \bs{\theta}_{i}, \bs{\beta}_{i}, \bs{Z}, \bs{y}_{i}$ from
        \begin{equation*}
            \mathcal{N}_T\left(\sigma_{i}^{-2}\bs{V}_{\bs{w}_i}(\bs{y}_i - \bs{Z}\bs{\beta}_i), \bs{V}_{\bs{w}_i} \right),
        \end{equation*}
        \Statex where $\bs{V}_{\bs{w}_i} = \left(\frac{1}{\sigma_{i}^{2}}\bs{I}  + \bs{R}(\bs{\theta}_i)^{-1}\right)^{-1}$
        \State update, for $i=1,\dots,n$, $\sigma_{i}^{2} \mid a_\sigma, b_\sigma, \bs{\beta}_i, \bs{w}_i, \bs{Z}, \bs{y}_{i}$ from
        \begin{equation*}
            \mbox{IG}\left( a_\sigma + \frac{T}{2}, b_\sigma + \frac{1}{2} (\bs{y}_{i} - \bs{Z}\bs{\beta}_i - \bs{w}_i)^\top(\bs{y}_{i} - \bs{Z}\bs{\beta}_i - \bs{w}_i) \right)
        \end{equation*}
        \State update, for $l=1,\dots,p$, $\zeta_{l}^{2} \mid a_\zeta, b_\zeta, \bs{\beta}_1,\dots,\bs{\beta}_n$ from
        \begin{equation*}
            \mbox{IG}\left( a_\zeta + \frac{n}{2}, b_\zeta + \frac{1}{2} \sum_{i=1}^n \beta_{il}^{2}\right)
        \end{equation*}
        \For{$i=1,\dots,n$}
        \State update $(\bs{\theta}_{1}^{\ast},\dots,\bs{\theta}_{K^{(-i)}}^{\ast})$ with the $K^{(-i)}$ unique values in $\bs{\theta}^{(-i)}$ 
        \State update $(\bs{\theta}_{K^{(-i)} + 1}^{\ast},\dots,\bs{\theta}_{K^{(-i)} + K_{aux}}^{\ast})$ from
        \begin{equation*}
            P_{0} = \mbox{IG}\left(a_{\tau},b_{\tau}\right) \times \mbox{Beta}\left(a_{\phi},b_{\phi}\right)
        \end{equation*}
        \State update $c_i$ with the following probability
        \begin{equation*}
            \operatorname{Pr}\left( c_{i} = k \mid - \right) \propto
            \begin{cases}
                f(\bs{w}_{i} \mid \bs{\theta}_{k}^{\ast}) \frac{C\left(S_{k}^{(-i)} \cup \{i\}\right)g\left(s_{k}^{{\ast}(-i)} \cup \bs{s}_{i} \right)}{C\left(S_{k}^{(-i)}\right)g\left(s_{k}^{{\ast}(-i)}\right)} & k = 1,\dots,K^{(-i)}, \\
                f(\bs{w}_{i} \mid \bs{\theta}_{k}^{\ast}) C\left(\{i\}\right)g\left(\bs{s}_{i}\right) &  k = K^{(-i)} + 1,\dots,K^{(-i)} + K_{aux},
            \end{cases}
        \end{equation*}
        \Statex where $f(\bs{w}_{i} \mid \bs{\theta}_{k}^{\ast})$ is the probability density function of $\bs{w}_{i}$ evaluated at the $k$-th cluster specific parameters, and $K_{aux}$ is the additional number of clusters; see \citet{neal2000markov}. 
        \State set $\bs{\theta}_{i} = \bs{\theta}_{c_{i}}^{\ast}$
        \EndFor
        \State update, for $k = 1,\dots, K$, $\phi_k^\ast \mid a_\phi, b_\phi, \bs{w}_1, \dots,\bs{w}_n, \bs{\theta}_k^\ast$ through a MH step and set, for $i=1,\dots,n$, $\phi_i = \phi_{c_i}^\ast$
        \State update, for $k = 1,\dots, K$, ${\tau_k^2}^\ast \mid a_\tau, b_\tau, \bs{w}_1, \dots,\bs{w}_n ,n_k, \phi_k^\ast$, from
        \begin{equation*}
            \mbox{IG}\left(a_\tau + \frac{n_k T}{2}, b_\tau + \frac{\sum_{i:c_i=k}\bs{w}_i^\top \Phi^{-1}(\phi_k^\ast) \bs{w}_i}{2} \right),
        \end{equation*}
        where $\Phi^{-1}(\phi_k^\ast) = {\tau_k^2}^\ast R(\bs{\theta}_k^\ast)^{-1}$, and set, for $i=1,\dots,n$, $\tau_i^2 = {\tau_{c_i}^2}^\ast$
        \If{$g(\cdot) \propto 1$}
        \State update $\alpha$
        \begin{align*}
            \alpha \mid \eta, K \sim &\frac{a_{\alpha} + K - 1}{a_{\alpha} + K - 1 + n(b_{\alpha} - \log(\eta))}\text{Gamma}(a_{\alpha} + K, b_{\alpha} - \log(\eta)) \\
            &+ \frac{n(b_{\alpha} - \log(\eta))}{a_{\alpha} + K - 1 + n(b_{\alpha} - \log(\eta))}\text{Gamma}(a_{\alpha} + K - 1, b_{\alpha} - \log(\eta)), 
        \end{align*}
        \Statex where $\eta \mid \alpha \sim \mbox{Gamma}(\alpha + 1, n)$.
        \EndIf
    \end{algorithmic}
\end{algorithm}

\section{Analysis of air quality data} \label{sec:data_analysis}

According to the World Health Organization, approximately 7 million premature deaths each year can be attributed to the exposure to indoor and outdoor air pollution \citep{who2022world}. Particulate matter (PM) is a standard proxy indicator for air pollution. Unlike other pollutants, PM is not a single chemical substance, but a complex mixture of solid and liquid particles that remain suspended in the atmosphere for extended periods, allowing them to undergo diffusion and transport processes. These particles originate from natural sources, such as soil erosion, volcanic activity, and pollen dispersal, as well as human activities, including industrial production, heating, and vehicular traffic. Exposure to PM poses serious health risks to living organisms and strongly affects the environment, contributing to climate change and contaminating soil and water. Therefore, continuous and widespread monitoring of PM levels is essential to prevent concentrations that are considered harmful to health. PM levels are typically measured at given time intervals by monitoring stations distributed across the region of interest. Hence, PM data are inherently spatio-temporal. In this section, we will illustrate how the Bayesian nonparametric clustering methodology illustrated in the previous sections can be applied to this type of data.

In the Bayesian literature, several approaches have been proposed to analyze PM data. For instance, \cite{sahu2006} introduced a Bayesian model with two random effects for rural and urban PM levels, while \cite{cocchi2007} proposed a Bayesian spatio-temporal model to predict daily PM concentrations. \cite{cameletti2011comparing} compared different Bayesian models for PM data based on goodness of fit and interpretability. \cite{hamm2015} presented a model with spatially varying coefficients for mapping PM levels over a large spatial domain. Finally, though this is by no means an exhaustive list, \cite{datta2016} introduced a Bayesian model based on an approximation of spatio-temporal Gaussian processes to analyze and forecast large-scale PM data.

\subsection{Data description}
\begin{figure}[tb]
    \centering
    \includegraphics[width=\textwidth]{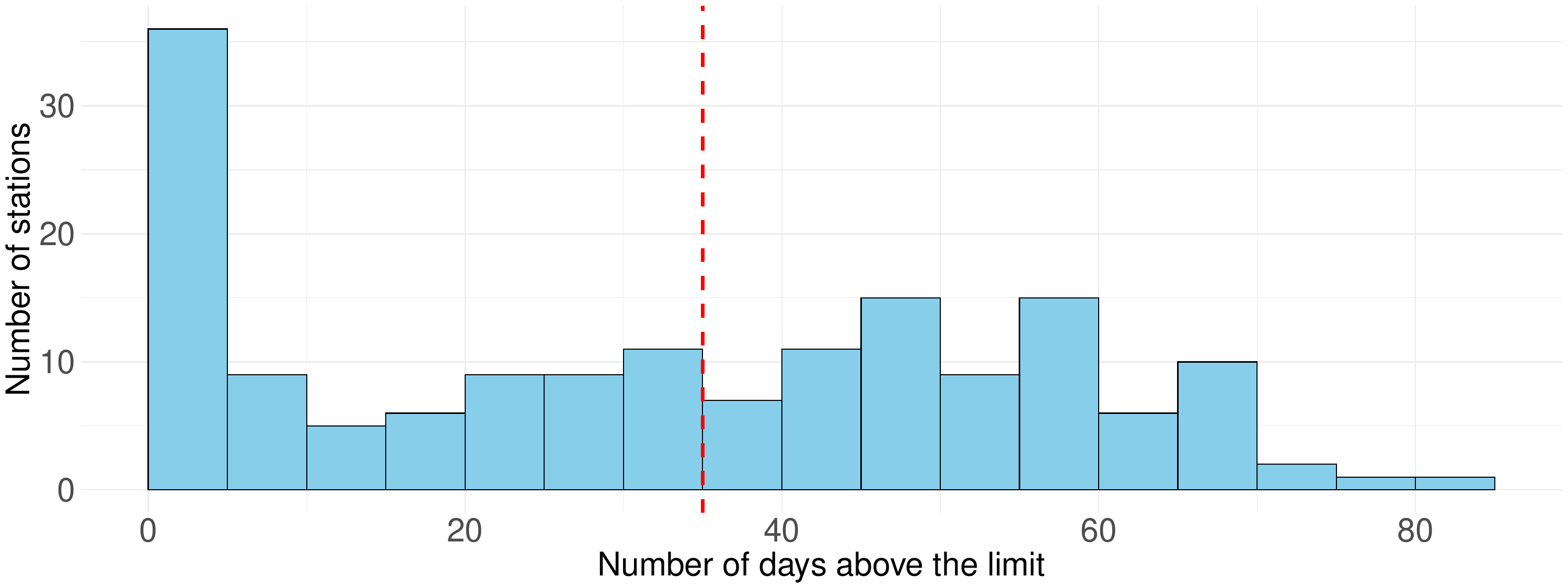}
    \caption{Monitoring stations above the PM10 daily limit in Northern Italy in 2019.}
    \label{fig:days_above_limit}
\end{figure}

In this work, we focus on a dataset consisting of daily PM10 concentration levels (PM with a diameter smaller than $10\mu m$), measured by 162 monitoring stations in Northern Italy from January 1st to December 31, 2019. The dataset was collected by the European Environment Agency (EEA) and is freely available at \url{https://eeadmz1-downloads-webapp.azurewebsites.net/}.   Figure~\ref{fig:days_above_limit} displays how many of these monitoring stations exceeded 50 $\mu g/m^{3}$, and for how many days. Note that, as established by the EEA \citep{eu2008directive}, each station should not exceed the concentration limit above on more than 35 days per year. However, in 2019, 77 out of the 162 stations in Northern Italy exceeded the limit on more than 35 days. In some particularly affected areas, stations registered even more than 80 days above the threshold. This underscores that air quality in Northern Italy is particularly critical and deserves careful monitoring. Air pollution in this area -- and in particular in the Po valley -- is a well-documented phenomenon, attributable to the high density of industrial activities, urban areas, and intensive livestock farming in a flat basin surrounded by mountains, which hinders effective air exchange \citep[e.g.,][]{larsen2012sources}.

\begin{figure}[b!]
    \centering
    \includegraphics[width =\linewidth]{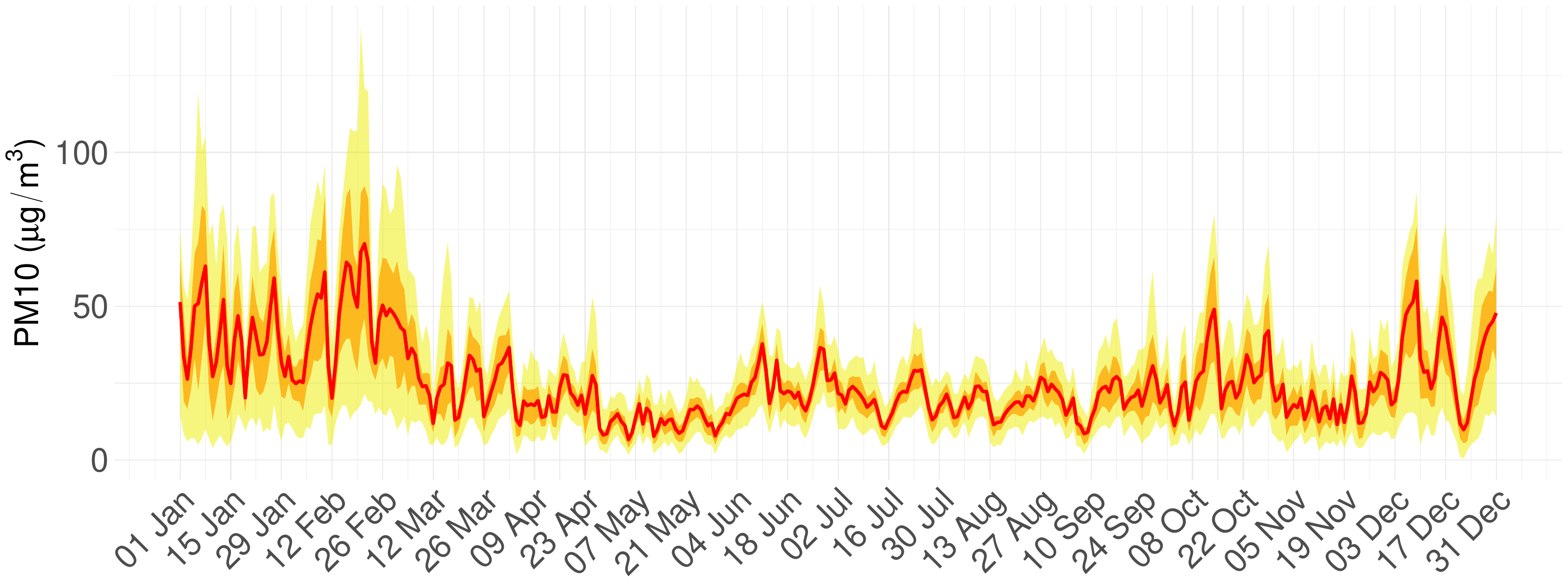} 
    \caption{Time series of the PM10 concentrations at the 162 considered monitoring stations in Northern Italy along the year 2019. The average across stations is reported in red, while interquartile and 90\% bands are reported in orange and yellow, respectively.}
    \label{fig:10ts}
\end{figure}

Figure~\ref{fig:10ts} presents the time series of the average PM10 concentrations across the considered monitoring stations over the 2019 -- along with its interquartile and 90\%  bands -- illustrating both the overall levels of pollution and its seasonal patterns. In particular, PM10 concentrations peak during colder months, with values frequently exceeding 50 $\mu g/m^{3}$ and occasionally approaching or even surpassing 100 $\mu g/m^{3}$. This seasonal pattern corresponds to known factors -- such as increased heating emissions, reduced atmospheric dispersion, and stagnant air conditions -- that are prevalent during winter \citep{pietrogrande2022seasonal}. In contrast, summer exhibits lower PM10 concentrations, often below 50 $\mu g/m^{3}$. The confidence bands in Figure~\ref{fig:10ts} also show considerable variability across stations.

\begin{figure}[b]
    \centering
    \begin{subfigure}{0.49\linewidth}
        \centering
        \includegraphics[width=\linewidth]{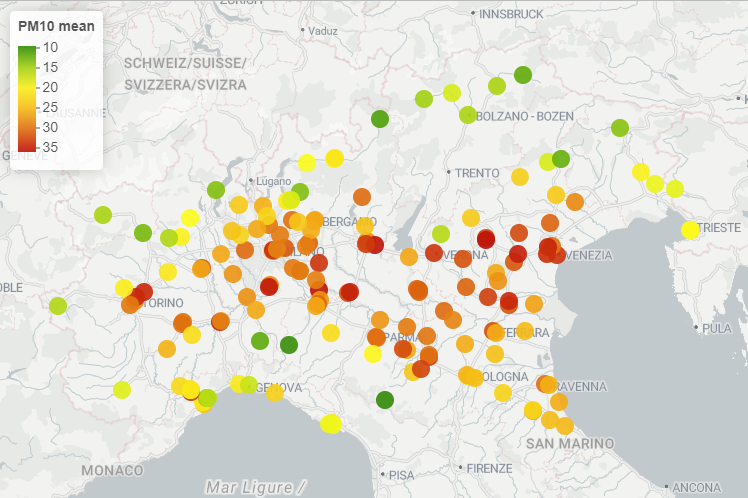}
        \caption{\label{fig:data_mean_map}}
    \end{subfigure}
    \hfill
    \begin{subfigure}{0.49\linewidth}
        \centering
        \includegraphics[width=\linewidth]{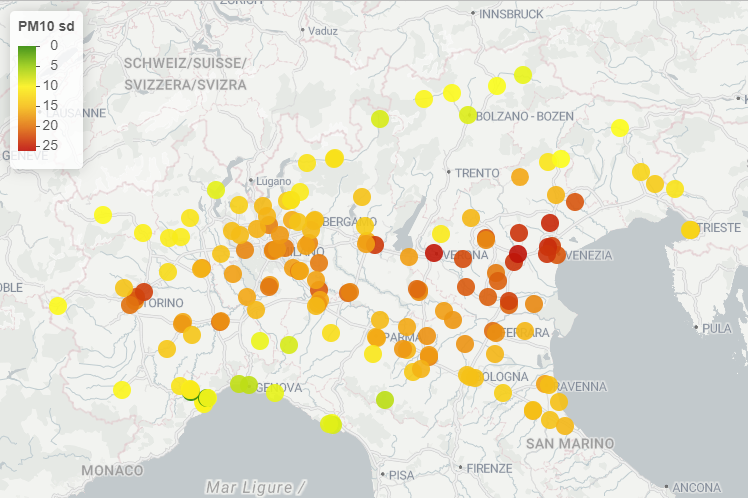}
        \caption{\label{fig:data_var_map}}
    \end{subfigure}
    \caption{Mean (\subref{fig:data_mean_map}) and standard deviation (\subref{fig:data_var_map}) of PM10 concentrations along year 2019 for each of the 162 considered monitoring
stations in Northern Italy.}
    \label{fig:pm10_data_maps}
\end{figure}

Instead, Figure~\ref{fig:pm10_data_maps} illustrates the mean and standard deviation of the PM10 concentration at each considered monitoring station. 
In particular, panel~(\subref{fig:data_mean_map}) shows that the highest mean values (often exceeding 30 $\mu g/m^{3}$) are recorded in urban and industrial areas like Turin, Milan, Brescia, Verona, and Venice, while lower mean concentrations (below 20 $\mu g/m^{3}$) are found in northern and coastal regions. 
Also note that the southern Po Valley exhibits moderate concentrations, generally between 20 and 30 $\mu g/m^{3}$.
Comparing panel~(\subref{fig:data_mean_map}) with panel~(\subref{fig:data_var_map}), which shows the standard deviation at each station, we can observe that regions with higher mean concentrations, particularly urban and industrial areas, also tend to experience greater temporal fluctuations. 

\subsection{Bayesian nonparametric clustering}\label{sec:data_analysis2}

In order to cluster monitoring stations, we adopt a dynamic model similar to the one described in \citet{nieto2014bayesian}. Specifically, the PM10 concentration recorded on day~$t$ at the monitoring station $i$, located at longitude and latitude $\bs{s}_{i}=(lon_{i},lat_{i})^\top$, is modeled as follows:
\begin{equation}\label{eq:data_analysis_mod}
    \begin{aligned}
        y_{t}(\bs{s}_{i}) &= \bs{z}_{t}^\top \bs{\beta}_{i} + w_{t}(\bs{s}_{i}) + \epsilon_{t}(\bs{s}_{i}), \quad &\text{with} \quad &\epsilon_{t}(\bs{s}_{i}) \overset{ind}{\sim} \mathcal{N}\left(0,\sigma_{i}^2\right),\\
        w_{t}(\bs{s}_{i}) &= \phi_{i} w_{t-1}(\bs{s}_{i}) + \nu_{t}(\bs{s}_{i}), \quad &\text{with} \quad &\nu_{t}(\bs{s}_{i}) \overset{ind}{\sim} \mathcal{N}\left(0,\tau_{i}^2\right),
    \end{aligned}
\end{equation}
for $i=1,\dots,n$ and $t=1,\dots,T$. Note that, differently from~\eqref{eq:gen_form}, the model in \eqref{eq:data_analysis_mod} does not include station-specific covariates. In place of the latter, it incorporates the vector $\bs{z}_t$,  which is the same across stations and comprises $p = 4$ dummy variables representing the intercept and seasonal effects. Specifically, $\bs{z}_t = (1,0,0,0)^\top$ for winter, $\bs{z}_t = (1,1,0,0)^\top$ for spring, $\bs{z}_t = (1,0,1,0)^\top$ for summer, and $\bs{z}_t = (1,0,0,1)^\top$ for autumn. 
The vector $\bs{\beta}_i$ instead comprises the $p$ location-specific regression coefficients, including the intercept. 

From the second line of \eqref{eq:data_analysis_mod}, it follows that the joint distribution of the spatio-temporal random effects $\bs{w}(\bs{s}_{i}) = (w_1(\bs{s}_i), \ldots, w_T(\bs{s}_i))^\top$ is
\begin{equation*}
    \bs{w}(\bs{s}_{i}) \sim \mathcal{N}_{T}\left(\bs{0},\bs{R}\left(\bs{\theta}_{i}\right)\right) \quad \text{with} \quad \left[\bs{R}\left(\bs{\theta}_{i}\right)\right]_{t,t^\prime} = \frac{\tau_{i}^2}{1-\phi_{i}^{2}} \phi_{i}^{\left|t-t^\prime\right|}, 
\end{equation*}
where $\bs{\theta}_{i} = \left( \phi_{i}, \tau_{i}^{2} \right)$. The clustering of the monitoring stations is then based on the parameters driving the autoregressive process of the random effect $w_t(\bs{s})$. Specifically, we employ the sPPM described in Section~\ref{sec:sPPM} to group stations according to the values of the parameters $\bs{\theta}_{i}$. In other words, we will group together monitoring stations whose time series have the same persistence and variability. This approach differs from clustering the entire time series, as discussed in Section~\ref{sec:spacetime}. Rather, we allow different time series to be grouped in the same cluster when they share the same underlying temporal process parameters.

After recalling that -- given a partition $\rho_n$ of the set $\{1, \dots, n\}$ and the cluster-specific parameters $\bs{\theta}_1^\ast, \dots, \bs{\theta}_K^\ast$ -- the parameter $\bs{\theta}_i$ is equal to $\bs{\theta}_k^\ast$ if $i \in S_k$ (see Section~\ref{sec:bayesian_clustering}), we specify a Bayesian clustering model by assigning the following prior distributions:
\begin{equation}
\label{eq:priors}
    \begin{aligned}
        \bs{\beta}_{i} &\overset{iid}{\sim} \mathcal{N}_{p}\left( \bs{0},  \text{diag}\left(\zeta_{1}^{2},\dots,\zeta_{p}^{2}\right) \right) &&\text{for} \quad i=1,\dots,n  \\ 
        \zeta_{l}^{2} &\overset{iid}{\sim} \text{IG}\left(a_\zeta,b_\zeta\right) &&\text{for} \quad l=1,\dots,p, \\ 
        \sigma_{i}^2 &\overset{iid}{\sim} \text{IG}\left(a_\sigma,b_\sigma\right) &&\text{for} \quad i=1,\dots,n,  \\
        \bs{\theta}_{k}^{\ast} \mid \rho_n &\overset{iid}{\sim} P_{0}, \quad P_0= \text{Beta}\left(a_{\phi},b_{\phi}\right) \times \text{IG}\left(a_{\tau},b_{\tau}\right) &&\text{for} \quad k = 1,\dots,K, \\
        p(\rho_n) &\propto \prod_{k=1}^{K} g(s_k^{\ast}) C\left(S_k\right), \\
    \end{aligned}
\end{equation}
where $\text{IG}\left(a,b\right)$ denotes an inverse gamma distribution with mean $b/(a - 1)$, while $\text{Beta}\left(a,b\right)$ denotes a beta distribution with mean $a/(a + b)$. 

In order to fully specify the prior on $\rho_n$, we must define both the cohesion function $C(S_k)$ and the similarity function $g(s_k^{\ast})$, as discussed in Section~\ref{sec:sPPM}. For the cohesion, we adopt the one associated with the DP, namely $C(S_k)=\alpha \Gamma(n_k)$. Regarding the similarity function, we consider two alternatives: in the first, we set $g(s_k^\ast) \propto 1$, corresponding to a standard PPM with no spatial covariate information. In this case, we add a hierarchy level to the model by setting $\alpha\sim\mbox {Gamma}(a_{\alpha}, b_{\alpha})$. In the second case, we employ the similarity in \eqref{eq:similarity3} -- thereby incorporating spatial information into the model -- with the popular Normal-Inverse-Wishart specification of $ q\left(\bs{\xi}_{k}\right)$. Note that, in this case, $\alpha$ is assumed to be fixed, since it appears in the normalizing constant of the EPPF, which cannot be computed in closed form or within a reasonable computational time.

Regarding the hyperparameters in \eqref{eq:priors}, we set $a_\zeta=a_\sigma=2$ and $b_\zeta=b_\sigma=1$ in order to induce a rather vague prior distribution (with mean one and infinite variance) on both the $\zeta_l^2$'s and the $\sigma_i^2$'s. This is a reasonable assumption in many practical applications where the variance of the residuals is expected to be of the same order of magnitude as that of the regression coefficients. As for the hyperparameters of the base measure $P_{0}$, we set $a_\phi=b_\phi=1$, thus inducing a Uniform(0,1) distribution on the AR(1) coefficient $\phi$. This imposes stationarity and positive autocorrelation on the univariate temporal process, which is consistent with our prior belief. Moreover, we set again $a_\tau=2$ and $b_\tau=1$ to allow for potentially large variances in the autoregressive process. Finally -- in the case of the PPM without covariates -- we set $a_\alpha = 2$ and $b_\alpha = 0.5$, thus leading to a fairly diffuse hyperprior on the DP parameter $\alpha$.

Posterior inference is then carried out through the Gibbs sampler detailed in Algorithm~\ref{alg:sppm_gs}. We finally employ the greedy stochastic search approach of \citet{dahl2022search}, based on the VI loss function, to obtain a posterior point estimate of the partition from the posterior samples.

\subsubsection{Results}

Here, we present the results of three alternative models: (i) a model without any clustering, i.e., $\bs{\theta}_{i} \overset{iid}{\sim} \text{IG}\left(a_{\tau},b_{\tau}\right) \times \text{Beta}\left(a_{\phi},b_{\phi}\right)$ for $i = 1,\dots,n$; (ii) a PPM without similarity function, and (iii) a sPPM based on the similarity function in~\eqref{eq:similarity3}. For each model, Algorithm~\ref{alg:sppm_gs} was run for 15,000 iterations, discarding the first 10,000 as burn-in. Note that in case (i) each observation was treated as a singleton cluster, while in case (iii) we set $\alpha$ equal to its posterior mean obtained in case (ii). The computational time per iteration was approximately 0.53 seconds for methods (ii)--(iii) and 0.72 seconds for method (i), i.e., the one without clustering. Further details and strategies to address computational bottlenecks are provided in the Appendices, Section~\ref{app:comp_det}. The \texttt{R} code to reproduce the results -- with the samplers implemented in \texttt{RcppArmadillo} \citep{armadillo2014} -- is available at \url{https://github.com/lucaaiello/time_series_clustering}.

\begin{figure}[tb]
    \centering
    \begin{subfigure}{0.49\linewidth}
        \centering
        \includegraphics[width=\linewidth]{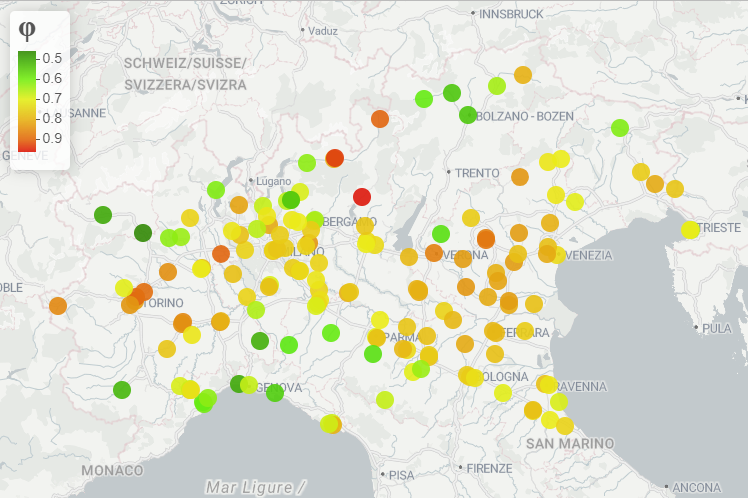}
        \caption{\label{fig:phi_map}}
    \end{subfigure}
    \hfill
    \begin{subfigure}{0.49\linewidth}
        \centering
        \includegraphics[width=\linewidth]{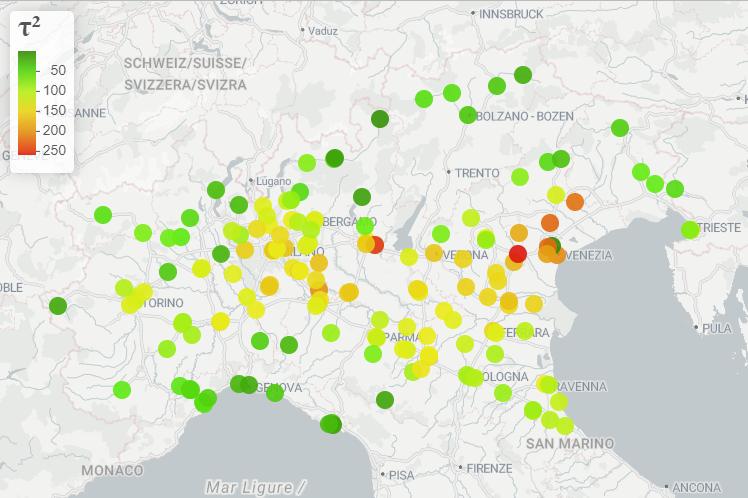}
        \caption{\label{fig:tau2_map}}
    \end{subfigure}
    \caption{Posterior means of the autoregressive coefficient $\phi_i$ (left panel) and the variance parameter $\tau^2_i$ (right panel) under the model without clustering.}
    \label{fig:noclust_map}
\end{figure}

Figure~\ref{fig:noclust_map} illustrates the posterior means of $\phi_{i}$ and $\tau_{i}^{2}$ from the model without clustering. In particular, panel (\subref{fig:phi_map}) displays the estimated $\phi_{i}$, with higher values (shown in red and orange) indicating strong persistence -- meaning that PM10 concentrations remain relatively stable over time. Panel (\subref{fig:tau2_map}) instead represents the estimated variability $\tau_{i}^{2}$ of PM10 levels. Notably, both maps reveal a spatial pattern, with stations that are geographically close tending to exhibit similar parameter values. In particular, the Po Valley exhibits both high persistence and high variability in PM10 concentrations, likely influenced by a combination of emissions, meteorological patterns, and geographical features that favor pollutant accumulation. 

\begin{figure}[b!]
    \centering
    \begin{subfigure}{0.39\linewidth}
        \centering
        \includegraphics[width=\linewidth]{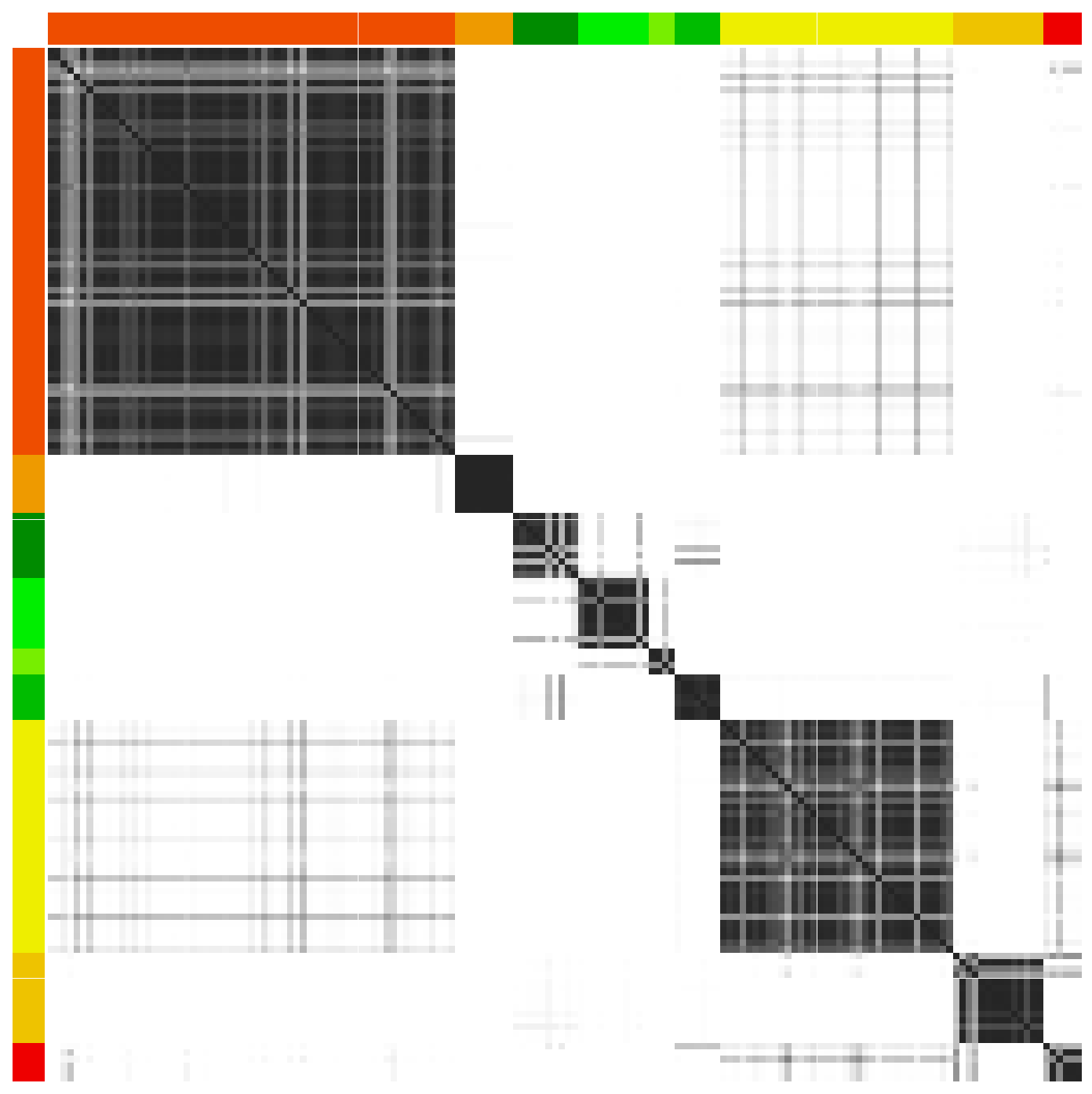}
        \caption{\label{fig:coclust_mat_cohesion}}
    \end{subfigure}
    \hfill
    \begin{subfigure}{0.59\linewidth}
        \centering
        \includegraphics[width=\linewidth]{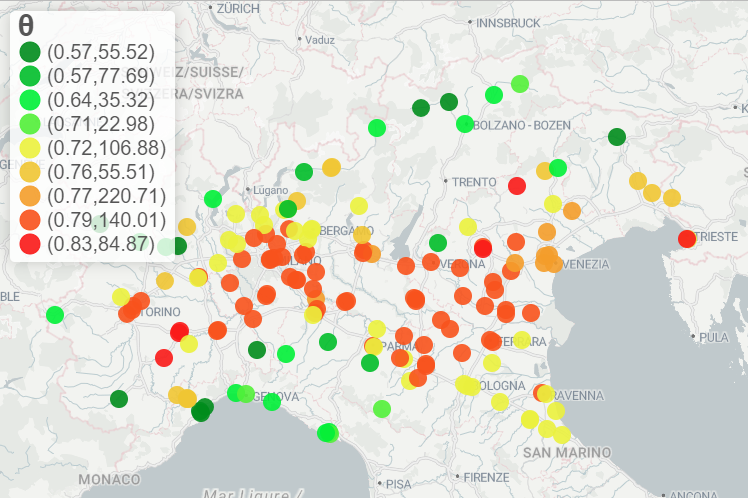}
        \caption{\label{fig:clust_map_cohesion}}
    \end{subfigure}
    \caption{Posterior co-clustering matrix (\subref{fig:coclust_mat_cohesion}) and monitoring stations colored according to the partition obtained from the PPM and the VI criterion (\subref{fig:clust_map_cohesion}). The legend displays the estimated cluster-specific values of $\bs{\theta}=(\phi,\tau^2)$.}
    
    \label{fig:cohesion_plots}
\end{figure}

Figure~\ref{fig:cohesion_plots} illustrates the results obtained using the PPM. In particular, panel (\subref{fig:coclust_mat_cohesion}) shows the estimated posterior probability that any two locations belong to the same cluster, with darker entries indicating higher co-clustering frequencies within posterior samples. Overall, this figure reveals low posterior uncertainty about the clustering structure. The VI loss yields a posterior point estimate of the partition consisting of nine clusters. Panel (\subref{fig:clust_map_cohesion}) shows the cluster assignment of the considered stations, and the corresponding cluster-specific values of $\bs{\theta}=(\phi,\tau^2)$. Specifically, the latter are estimated by running Algorithm~\ref{alg:sppm_gs}, conditioning on the partition estimated using the VI loss function, as described in Section~\ref{subsec:post_inf}. The color gradient helps to highlight the spatial patterns, revealing higher persistence and variability in central areas of the Po Valley, while lower values are mainly found in coastal and mountainous areas. 

\begin{figure}[tb]
    \centering
    \begin{subfigure}{0.39\linewidth}
        \centering
        \includegraphics[width=\linewidth]{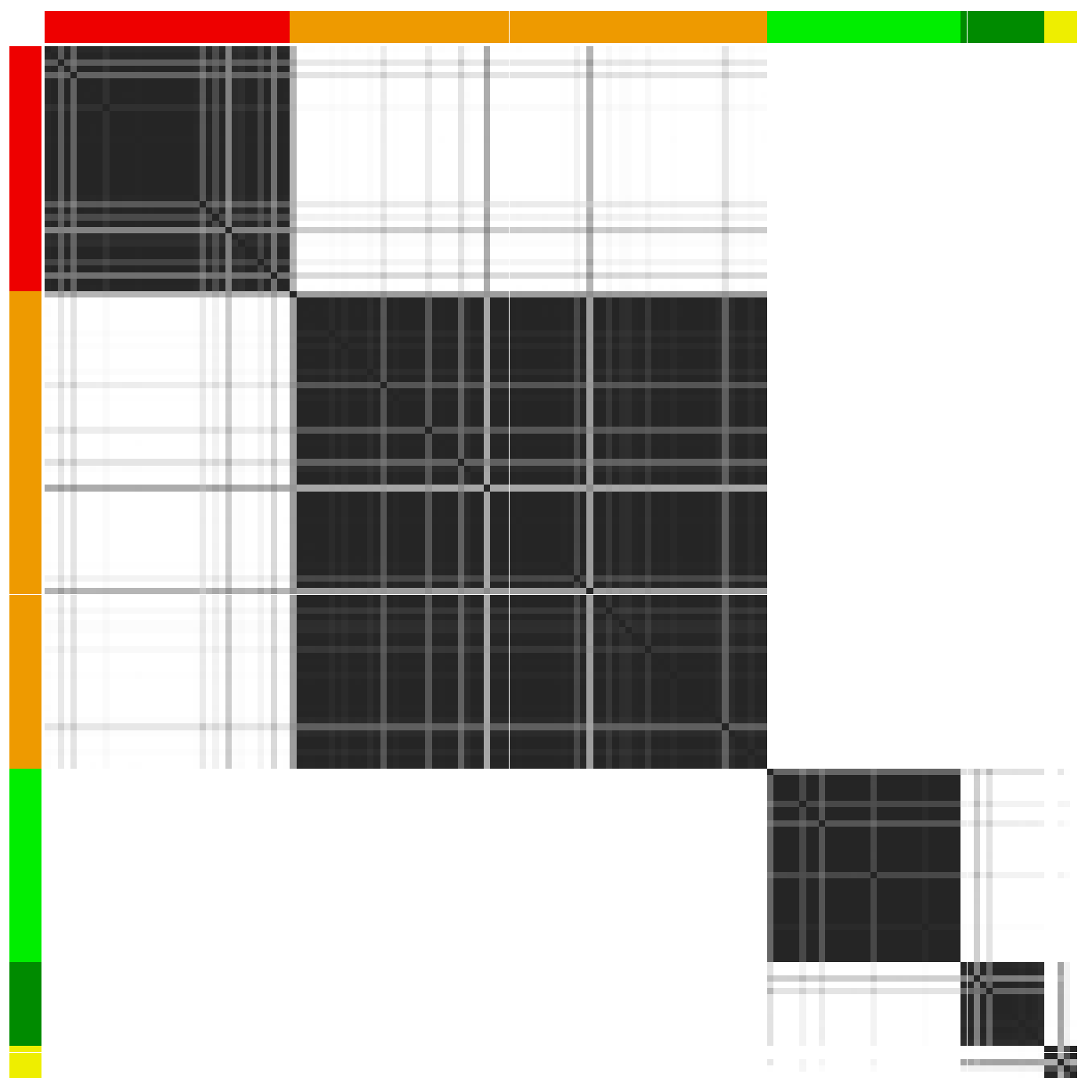}
        \caption{\label{fig:coclust_mat_similarity}}
    \end{subfigure}
    \hfill
    \begin{subfigure}{0.59\linewidth}
        \centering
        \includegraphics[width=\linewidth]{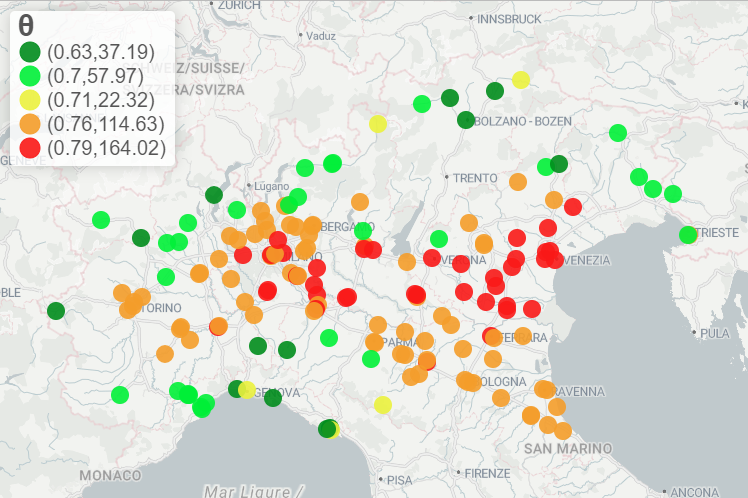}
        \caption{\label{fig:clust_map_similarity}}
    \end{subfigure}
    \caption{Posterior co-clustering matrix (\subref{fig:coclust_mat_cohesion}) and monitoring stations colored according to the partition obtained from the sPPM and the VI criterion (\subref{fig:clust_map_cohesion}). The legend displays the estimated cluster-specific values of $\bs{\theta}=(\phi,\tau^2)$.}
    \label{fig:similarity_plots}
\end{figure}

The results obtained after incorporating the similarity function into the clustering procedure -- i.e., under the sPPM -- are instead presented in Figure~\ref{fig:similarity_plots}.  The co-clustering matrix in panel~(\subref{fig:coclust_mat_similarity}) reflects a further reduction of uncertainty about the partition with respect to the PPM without spatial covariates -- i.e., with respect to panel~(\subref{fig:coclust_mat_cohesion}) of Figure~\ref{fig:cohesion_plots} -- with both a smaller number of clusters and lower uncertainty in assigning stations to clusters. Panel~(\subref{fig:clust_map_similarity}) displays the monitoring stations colored according to the estimated clustering and illustrates how the introduction of the similarity function has favored the co-clustering of monitoring stations that were previously assigned to different clusters despite being close to each other. The resulting map reveals three distinct regions: the central Po Valley (red cluster), the surrounding area adjacent to the Alps and the Apennines (orange cluster), and the coastal and mountainous areas (yellow to green clusters) -- with decreasing persistence and variability (with a single exception). 

\begin{figure}[bt]
    \centering
    \includegraphics[width=.8\linewidth]{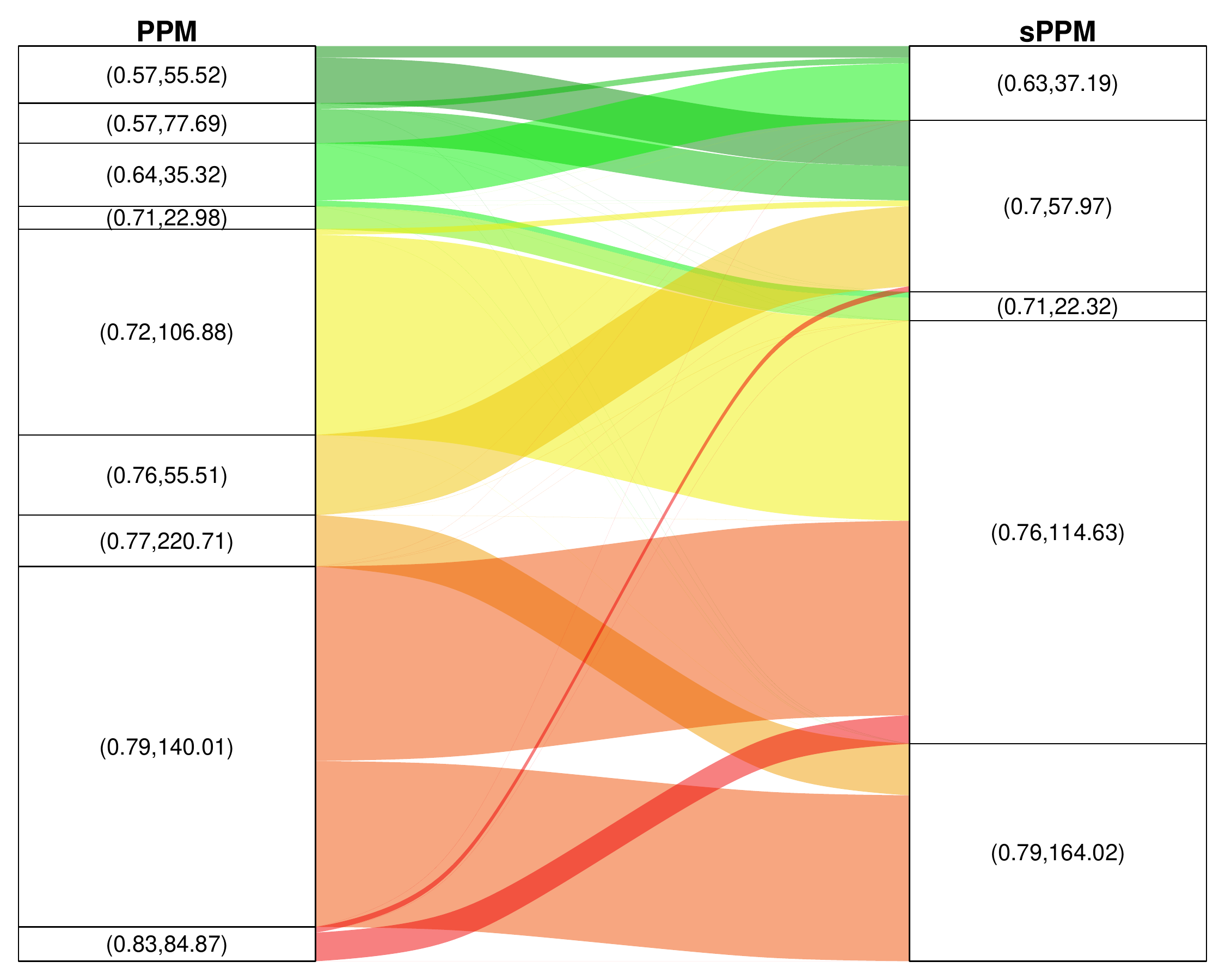}
    \caption{Riverplot illustrating the difference between the VI-loss point estimates of the partitions obtained from the PPM (left) and the sPPM (right). The values in the boxes display the estimated luster-specific parameters 
    $\bs{\theta}^\ast_{k} = (\phi^\ast_{k}, {\tau^2_k}^\ast)$ 
    under the PPM and the sPPM, respectively.}
    \label{fig:riverplot}
\end{figure}

Finally, the riverplot in Figure~\ref{fig:riverplot} further illustrates the difference between the VI-loss point estimates of the partitions obtained from the PPM (left) and the sPPM (right).  Colors refer to the clusters obtained under the PPM, while numbers reported within the boxes are the estimated cluster-specific parameters, computed by running Algorithm~\ref{alg:sppm_gs} conditioned on the estimated partition. Note, again, that leveraging spatial information
yields fewer clusters (five instead of nine), thus yielding a more parsimonious and interpretable clustering structure.

To further illustrate the results of our clustering method, in the Appendices (Section~\ref{app:additional}) we present the time series means along with interquartile and 90\% bands for each of the clusters obtained from both PPM and sPPM. These plots confirm that clusters exhibiting higher persistence and variability correspond to areas experiencing more severe PM10 levels, whereas those with lower persistence and variability tend to record more moderate PM10 concentrations.

\section{Summary and future directions}\label{sec:discussion}

In this work, we provided an overview of Bayesian nonparametric methods for clustering spatio-temporal environmental data. We first reviewed models for spatio-temporal data, and then discussed Bayesian nonparametric methods for clustering, also showing how to include individual covariates (e.g., geographical coordinates). We then specialized this approach to spatio-temporal data, which are ubiquitous in environmental science. Finally, we illustrated the proposed methods on PM10 monitoring data from the Po Valley in Northern Italy, demonstrating their ability to identify meaningful spatial patterns.

In particular, in the application presented in this work, we employed a multi-level hierarchical linear mixture model, incorporating a first-order autoregressive process to account for temporal effects, and clustering together time series exhibiting the same autoregressive coefficient and temporal variance. We then enriched our nonparametric model to induce spatial homogeneity on the clustering by including geographical coordinates via a sPPM framework. 

Future research in Bayesian nonparametric clustering for spatio-temporal environmental data could focus on exploring alternative cohesion and similarity functions within the sPPM. Additionally, extending the sPPM approach to handle areal data would represent an original contribution of substantial practical relevance. Another important research direction may be the refinement of hyperparameter selection to enhance model performance. Finally, a gap appears in the literature regarding the predictive capabilities of these models. Although they have shown strong inferential properties, limited research has addressed their predictive performance, which is crucial for understanding environmental changes.

\bibliographystyle{apalike}
\bibliography{biblio}

\begin{thebibliography}{}

\bibitem[Argiento and De~Iorio, 2022]{argiento2022infinity}
Argiento, R. and De~Iorio, M. (2022).
\newblock Is infinity that far? {A} {B}ayesian nonparametric perspective of finite mixture models.
\newblock {\em The Annals of Statistics}, 50(5):2641--2663.

\bibitem[Argiento et~al., 2024]{argiento2024}
Argiento, R., Filippi-Mazzola, E., and Paci, L. (2024).
\newblock Model-based clustering of categorical data based on the {H}amming distance.
\newblock {\em Journal of the American Statistical Association}, doi:10.1080/01621459.2024.2402568:1--20.

\bibitem[Banerjee et~al., 2014]{banerjee2014hierarchical}
Banerjee, S., Carlin, B.~P., and Gelfand, A.~E. (2014).
\newblock {\em Hierarchical Modeling and Analysis for Spatial Data}.
\newblock CRC Press.

\bibitem[Beraha et~al., 2025]{beraha2025}
Beraha, M., Argiento, R., Camerlenghi, F., and Guglielmi, A. (2025).
\newblock Bayesian mixtures models with repulsive and attractive atoms.
\newblock {\em arXiv 2302.09034}.

\bibitem[Berrocal, 2016]{berrocal2016identifying}
Berrocal, V. (2016).
\newblock Identifying trends in the spatial errors of a regional climate model via clustering.
\newblock {\em Environmetrics}, 27(2):90--102.

\bibitem[Bezdek et~al., 1984]{bezdek1984fcm}
Bezdek, J.~C., Ehrlich, R., and Full, W. (1984).
\newblock {FCM: The fuzzy c-means clustering algorithm}.
\newblock {\em Computers \& geosciences}, 10(2-3):191--203.

\bibitem[Binder, 1978]{binder1978bayesian}
Binder, D.~A. (1978).
\newblock Bayesian cluster analysis.
\newblock {\em Biometrika}, 65(1):31--38.

\bibitem[Bitto and Fr{\"u}hwirth-Schnatter, 2019]{bitto2019achieving}
Bitto, A. and Fr{\"u}hwirth-Schnatter, S. (2019).
\newblock Achieving shrinkage in a time-varying parameter model framework.
\newblock {\em Journal of Econometrics}, 210(1):75--97.

\bibitem[Bouveyron et~al., 2019]{bouveyron2019model}
Bouveyron, C., Celeux, G., Murphy, T.~B., and Raftery, A.~E. (2019).
\newblock {\em Model-based clustering and classification for data science: with applications in R}, volume~50.
\newblock Cambridge University Press.

\bibitem[Bucci et~al., 2022]{bucci2022clustering}
Bucci, A., Ippoliti, L., Valentini, P., and Fontanella, S. (2022).
\newblock Clustering spatio-temporal series of confirmed {COVID}-19 deaths in europe.
\newblock {\em Spatial Statistics}, 49:100543.

\bibitem[Cameletti et~al., 2011]{cameletti2011comparing}
Cameletti, M., Ignaccolo, R., and Bande, S. (2011).
\newblock Comparing spatio-temporal models for particulate matter in {P}iemonte.
\newblock {\em Environmetrics}, 22(8):985--996.

\bibitem[Cheam et~al., 2017]{cheam2017model}
Cheam, A., Marbac, M., and McNicholas, P. (2017).
\newblock Model-based clustering for spatiotemporal data on air quality monitoring.
\newblock {\em Environmetrics}, 28(3):e2437.

\bibitem[Cocchi et~al., 2007]{cocchi2007}
Cocchi, D., Greco, F., and Trivisano, C. (2007).
\newblock Hierarchical space-time modelling of {PM10} pollution.
\newblock {\em Atmospheric Environment}, 41(3):532--542.

\bibitem[Cressie and Wikle, 2015]{cressie2015statistics}
Cressie, N. and Wikle, C.~K. (2015).
\newblock {\em Statistics for spatio-temporal data}.
\newblock John Wiley \& Sons.

\bibitem[Dahl et~al., 2022]{dahl2022search}
Dahl, D.~B., Johnson, D.~J., and M{\"u}ller, P. (2022).
\newblock Search algorithms and loss functions for {B}ayesian clustering.
\newblock {\em Journal of Computational and Graphical Statistics}, 31(4):1189--1201.

\bibitem[Dahl et~al., 2021]{salso}
Dahl, D.~B., Johnson, D.~J., and Müller, P. (2021).
\newblock {\em salso: Search Algorithms and Loss Functions for {B}ayesian Clustering}.
\newblock R package version 0.2.22.

\bibitem[Dasgupta and Raftery, 1998]{dasgupta1998detecting}
Dasgupta, A. and Raftery, A.~E. (1998).
\newblock Detecting features in spatial point processes with clutter via model-based clustering.
\newblock {\em Journal of the American statistical Association}, 93(441):294--302.

\bibitem[Datta et~al., 2016]{datta2016}
Datta, A., Banerjee, S., Finley, A.~O., Hamm, N. A.~S., and Schaap, M. (2016).
\newblock {Nonseparable dynamic nearest neighbor Gaussian process models for large spatio-temporal data with an application to particulate matter analysis}.
\newblock {\em The Annals of Applied Statistics}, 10(3):1286--1316.

\bibitem[De~Blasi et~al., 2015]{de2015gibbs}
De~Blasi, P., Favaro, S., Lijoi, A., Mena, R., Pr{\"u}nster, I., and Ruggiero, M. (2015).
\newblock Are {G}ibbs-type priors the most natural generalization of the {D}irichlet process?
\newblock {\em IEEE Transactions on Pattern Analysis and Machine Intelligence}, 37(2):212--229.

\bibitem[Duan et~al., 2007]{duan2007generalized}
Duan, J.~A., Guindani, M., and Gelfand, A.~E. (2007).
\newblock Generalized spatial {D}irichlet process models.
\newblock {\em Biometrika}, 94(4):809--825.

\bibitem[Eddelbuettel and Sanderson, 2014]{armadillo2014}
Eddelbuettel, D. and Sanderson, C. (2014).
\newblock {RcppArmadillo: Accelerating R with high-performance C++ linear algebra}.
\newblock {\em Computational Statistics and Data Analysis}, 71:1054--1063.

\bibitem[Escobar, 1994]{escobar1994estimating}
Escobar, M.~D. (1994).
\newblock Estimating {N}ormal means with a {D}irichlet process prior.
\newblock {\em Journal of the American Statistical Association}, 89(425):268--277.

\bibitem[Escobar and West, 1995]{escobar1995bayesian}
Escobar, M.~D. and West, M. (1995).
\newblock Bayesian density estimation and inference using mixtures.
\newblock {\em Journal of the American Statistical Association}, 90(430):577--588.

\bibitem[{European Commission}, 2008]{eu2008directive}
{European Commission} (2008).
\newblock Directive 2008/50/ec of the {European Parliament and of the Council of 21 May 2008 on ambient air quality and cleaner air for Europe}.
\newblock \url{https://eur-lex.europa.eu/eli/dir/2008/50/2015-09-18}.

\bibitem[Ferguson, 1973]{ferguson1973bayesian}
Ferguson, T.~S. (1973).
\newblock A {B}ayesian analysis of some nonparametric problems.
\newblock {\em The Annals of Statistics}, 1(2):209--230.

\bibitem[Fern{\'a}ndez and Green, 2002]{fernandez2002modelling}
Fern{\'a}ndez, C. and Green, P.~J. (2002).
\newblock Modelling spatially correlated data via mixtures: a {B}ayesian approach.
\newblock {\em Journal of the Royal Statistical Society Series B: Statistical Methodology}, 64(4):805--826.

\bibitem[Fraley and Raftery, 2002]{fraley2002model}
Fraley, C. and Raftery, A.~E. (2002).
\newblock Model-based clustering, discriminant analysis, and density estimation.
\newblock {\em Journal of the American statistical Association}, 97(458):611--631.

\bibitem[Franczak et~al., 2013]{franczak2013mixtures}
Franczak, B.~C., Browne, R.~P., and McNicholas, P.~D. (2013).
\newblock Mixtures of shifted asymmetric {L}aplace distributions.
\newblock {\em IEEE Transactions on Pattern Analysis and Machine Intelligence}, 36(6):1149--1157.

\bibitem[Fr{\"u}hwirth-Schnatter et~al., 2019]{fruhwirth2019handbook}
Fr{\"u}hwirth-Schnatter, S., Celeux, G., and Robert, C.~P. (2019).
\newblock {\em Handbook of mixture analysis}.
\newblock CRC press.

\bibitem[Fr{\"u}hwirth-Schnatter and Kaufmann, 2008]{fruhwirth2008model}
Fr{\"u}hwirth-Schnatter, S. and Kaufmann, S. (2008).
\newblock Model-based clustering of multiple time series.
\newblock {\em Journal of Business \& Economic Statistics}, 26(1):78--89.

\bibitem[Fr{\"u}hwirth-Schnatter et~al., 2021]{fruhwirth2021generalized}
Fr{\"u}hwirth-Schnatter, S., Malsiner-Walli, G., and Gr{\"u}n, B. (2021).
\newblock Generalized mixtures of finite mixtures and telescoping sampling.
\newblock {\em Bayesian Analysis}, 16(4):1279--1307.

\bibitem[Fr{\"u}hwirth-Schnatter and Pyne, 2010]{fruhwirth2010bayesian}
Fr{\"u}hwirth-Schnatter, S. and Pyne, S. (2010).
\newblock Bayesian inference for finite mixtures of univariate and multivariate skew-{N}ormal and skew-t distributions.
\newblock {\em Biostatistics}, 11(2):317--336.

\bibitem[F{\'u}quene et~al., 2019]{fuquene2019choosing}
F{\'u}quene, J., Steel, M., and Rossell, D. (2019).
\newblock On choosing mixture components via non-local priors.
\newblock {\em Journal of the Royal Statistical Society Series B: Statistical Methodology}, 81(5):809--837.

\bibitem[Gelfand et~al., 2005]{gelfand2005bayesian}
Gelfand, A.~E., Kottas, A., and MacEachern, S.~N. (2005).
\newblock Bayesian nonparametric spatial modeling with {D}irichlet process mixing.
\newblock {\em Journal of the American Statistical Association}, 100(471):1021--1035.

\bibitem[Gnedin, 2010]{gnedin2010species}
Gnedin, A. (2010).
\newblock A species sampling model with finitely many types.
\newblock {\em Electronic Communications in Probability [electronic only]}, 15:79--88.

\bibitem[Gnedin and Pitman, 2006]{gnedin2006exchangeable}
Gnedin, A. and Pitman, J. (2006).
\newblock Exchangeable {G}ibbs partitions and {S}tirling triangles.
\newblock {\em Journal of Mathematical sciences}, 138:5674--5685.

\bibitem[Goodman, 1974]{goodman1974exploratory}
Goodman, L.~A. (1974).
\newblock Exploratory latent structure analysis using both identifiable and unidentifiable models.
\newblock {\em Biometrika}, 61(2):215--231.

\bibitem[Grazian, 2023]{grazian2023review}
Grazian, C. (2023).
\newblock A review on {B}ayesian model-based clustering.
\newblock {\em arXiv preprint arXiv:2303.17182}.

\bibitem[Grazian et~al., 2020]{grazian2020loss}
Grazian, C., Villa, C., and Liseo, B. (2020).
\newblock On a loss-based prior for the number of components in mixture models.
\newblock {\em Statistics \& Probability Letters}, 158:108656.

\bibitem[Green, 1995]{green1995}
Green, P.~J. (1995).
\newblock Reversible jump {Markov chain Monte Carlo} computation and {B}ayesian model determination.
\newblock {\em Biometrika}, 82(4):711--732.

\bibitem[Griffiths et~al., 1974]{griffiths1974properties}
Griffiths, R.~C., Engen, G.~F., and McCloskey, R.~J. (1974).
\newblock Properties of the two-parameter {Poisson–Dirichlet} distribution.
\newblock {\em Journal of Applied Probability}, 11(2):319--338.

\bibitem[Hamm et~al., 2015]{hamm2015}
Hamm, N., Finley, A., Schaap, M., and Stein, A. (2015).
\newblock A spatially varying coefficient model for mapping {PM10} air quality at the {E}uropean scale.
\newblock {\em Atmospheric Environment}, 102:393 -- 405.

\bibitem[Harrison and Stevens, 1976]{harrison1976bayesian}
Harrison, P.~J. and Stevens, C.~F. (1976).
\newblock Bayesian forecasting.
\newblock {\em Journal of the Royal Statistical Society Series B: Statistical Methodology}, 38(3):205--228.

\bibitem[Hartigan, 1990]{hartigan1990partition}
Hartigan, J.~A. (1990).
\newblock Partition {M}odels.
\newblock {\em Communications in statistics-Theory and methods}, 19(8):2745--2756.

\bibitem[Hefley et~al., 2017]{hefley2017dynamic}
Hefley, T.~J., Hooten, M.~B., Hanks, E.~M., Russell, R.~E., and Walsh, D.~P. (2017).
\newblock Dynamic spatio-temporal models for spatial data.
\newblock {\em Spatial statistics}, 20:206--220.

\bibitem[Ishwaran and James, 2001]{ishwaran2001gibbs}
Ishwaran, H. and James, L.~F. (2001).
\newblock Gibbs sampling methods for stick-breaking priors.
\newblock {\em Journal of the American statistical Association}, 96(453):161--173.

\bibitem[Jasra et~al., 2005]{MCMC_label_switch}
Jasra, A., Holmes, C.~C., and Stephens, D.~A. (2005).
\newblock Markov chain {Monte Carlo} methods and the label switching problem in {B}ayesian mixture modeling.
\newblock {\em Statistical Science}, 20(1):50--67.

\bibitem[Kalli et~al., 2011]{kalli2011slice}
Kalli, M., Griffin, J.~E., and Walker, S.~G. (2011).
\newblock Slice sampling mixture models.
\newblock {\em Statistics and computing}, 21:93--105.

\bibitem[Karlis and Xekalaki, 2005]{karlis2005mixed}
Karlis, D. and Xekalaki, E. (2005).
\newblock Mixed {P}oisson distributions.
\newblock {\em International Statistical Review/Revue Internationale de Statistique}, pages 35--58.

\bibitem[Kastner and Fr{\"u}hwirth-Schnatter, 2014]{kastner2014ancillarity}
Kastner, G. and Fr{\"u}hwirth-Schnatter, S. (2014).
\newblock Ancillarity-sufficiency interweaving strategy ({ASIS}) for boosting mcmc estimation of stochastic volatility models.
\newblock {\em Computational Statistics \& Data Analysis}, 76:408--423.

\bibitem[Knorr-Held and Besag, 1998]{knorr1998modelling}
Knorr-Held, L. and Besag, J. (1998).
\newblock Modelling risk from a disease in time and space.
\newblock {\em Statistics in medicine}, 17(18):2045--2060.

\bibitem[Kottas et~al., 2008]{kottas2008modeling}
Kottas, A., Duan, J.~A., and Gelfand, A.~E. (2008).
\newblock Modeling disease incidence data with spatial and spatio temporal {D}irichlet process mixtures.
\newblock {\em Biometrical Journal: Journal of Mathematical Methods in Biosciences}, 50(1):29--42.

\bibitem[Krnjaji{\'c} et~al., 2008]{krnjajic2008parametric}
Krnjaji{\'c}, M., Kottas, A., and Draper, D. (2008).
\newblock Parametric and nonparametric {B}ayesian model specification: A case study involving models for count data.
\newblock {\em Computational Statistics \& Data Analysis}, 52(4):2110--2128.

\bibitem[Larsen et~al., 2012]{larsen2012sources}
Larsen, B., Gilardoni, S., Stenstr{\"o}m, K., Niedzialek, J., Jimenez, J., and Belis, C. (2012).
\newblock {Sources for PM air pollution in the Po Plain, Italy: II. Probabilistic uncertainty characterization and sensitivity analysis of secondary and primary sources}.
\newblock {\em Atmospheric Environment}, 50:203--213.

\bibitem[Laurini, 2019]{laurini2019spatio}
Laurini, M.~P. (2019).
\newblock A spatio-temporal approach to estimate patterns of climate change.
\newblock {\em Environmetrics}, 30(1):e2542.

\bibitem[Lee et~al., 2021]{lee2021clustered}
Lee, J., Kamenetsky, M.~E., Gangnon, R.~E., and Zhu, J. (2021).
\newblock Clustered spatio-temporal varying coefficient regression model.
\newblock {\em Statistics in medicine}, 40(2):465--480.

\bibitem[Lee and McLachlan, 2014]{lee2014finite}
Lee, S. and McLachlan, G.~J. (2014).
\newblock Finite mixtures of multivariate skew t-distributions: some recent and new results.
\newblock {\em Statistics and Computing}, 24:181--202.

\bibitem[Li et~al., 2016]{li2016comparison}
Li, B., Zhang, X., and Smerdon, J.~E. (2016).
\newblock Comparison between spatio-temporal random processes and application to climate model data.
\newblock {\em Environmetrics}, 27(5):267--279.

\bibitem[Lijoi et~al., 2008]{lijoi2008bayesian}
Lijoi, A., Prunster, I., and Walker, S.~G. (2008).
\newblock Bayesian nonparametric estimators derived from conditional {G}ibbs structures.
\newblock {\em Annals of Applied Probability}, 18(4):1519--1547.

\bibitem[Liu et~al., 2024]{liu2024shared}
Liu, J., Wade, S., and Bochkina, N. (2024).
\newblock Shared differential clustering across single-cell rna sequencing datasets with the hierarchical {D}irichlet process.
\newblock {\em Econometrics and Statistics}.

\bibitem[Lo, 1984]{lo1984class}
Lo, A.~Y. (1984).
\newblock On a class of {B}ayesian nonparametric estimates: I. density estimates.
\newblock {\em The Annals of Statistics}, pages 351--357.

\bibitem[MacEachern, 1994]{maceachern1994estimating}
MacEachern, S.~N. (1994).
\newblock Estimating {N}ormal means with a conjugate style {D}irichlet process prior.
\newblock {\em Communications in Statistics-Simulation and Computation}, 23(3):727--741.

\bibitem[MacEachern, 1998]{maceachern1998computational}
MacEachern, S.~N. (1998).
\newblock Computational methods for mixture of {D}irichlet process models.
\newblock In {\em Practical nonparametric and semiparametric Bayesian statistics}, pages 23--43. Springer.

\bibitem[MacEachern and M{\"u}ller, 1998]{maceachern1998estimating}
MacEachern, S.~N. and M{\"u}ller, P. (1998).
\newblock Estimating mixture of {D}irichlet process models.
\newblock {\em Journal of Computational and Graphical Statistics}, 7(2):223--238.

\bibitem[Mastrantonio et~al., 2019]{mastrantonio2019new}
Mastrantonio, G., Grazian, C., Mancinelli, S., and Bibbona, E. (2019).
\newblock New formulation of the logistic-{G}aussian process to analyze trajectory tracking data.
\newblock {\em The Annals of Applied Statistics}, 13(4):2483--2508.

\bibitem[McCausland et~al., 2011]{mccausland2011simulation}
McCausland, W.~J., Miller, S., and Pelletier, D. (2011).
\newblock Simulation smoothing for state--space models: A computational efficiency analysis.
\newblock {\em Computational Statistics \& Data Analysis}, 55(1):199--212.

\bibitem[Meil{\u{a}}, 2007]{meilua2007comparing}
Meil{\u{a}}, M. (2007).
\newblock Comparing clusterings--an information based distance.
\newblock {\em Journal of multivariate analysis}, 98(5):873--895.

\bibitem[Meurant, 1992]{meurant1992review}
Meurant, G. (1992).
\newblock A review on the inverse of symmetric tridiagonal and block tridiagonal matrices.
\newblock {\em SIAM Journal on Matrix Analysis and Applications}, 13(3):707--728.

\bibitem[Miller and Harrison, 2018]{miller2018mixture}
Miller, J.~W. and Harrison, M.~T. (2018).
\newblock Mixture models with a prior on the number of components.
\newblock {\em Journal of the American Statistical Association}, 113(521):340--356.

\bibitem[Molitor et~al., 2010]{molitor2010bayesian}
Molitor, J., Papathomas, M., Jerrett, M., and Richardson, S. (2010).
\newblock Bayesian profile regression with an application to the {National Survey of Children's Health}.
\newblock {\em Biostatistics}, 11(3):484--498.

\bibitem[M{\"u}ller et~al., 2011]{muller2011product}
M{\"u}ller, P., Quintana, F., and Rosner, G.~L. (2011).
\newblock A {P}roduct {P}artition {M}odel with regression on covariates.
\newblock {\em Journal of Computational and Graphical Statistics}, 20(1):260--278.

\bibitem[M{\"u}ller et~al., 2015]{muller2015bayesian}
M{\"u}ller, P., Quintana, F.~A., Jara, A., and Hanson, T. (2015).
\newblock {\em Bayesian nonparametric data analysis}, volume~1.
\newblock Springer.

\bibitem[Musau et~al., 2022]{musau2022clustering}
Musau, V.~M., Gaetan, C., and Girardi, P. (2022).
\newblock Clustering of bivariate satellite time series: A quantile approach.
\newblock {\em Environmetrics}, 33(7):e2755.

\bibitem[Neal, 2000]{neal2000markov}
Neal, R.~M. (2000).
\newblock Markov chain sampling methods for {D}irichlet process mixture models.
\newblock {\em Journal of computational and graphical statistics}, 9(2):249--265.

\bibitem[Nguyen and Gelfand, 2011]{nguyen2011dirichlet}
Nguyen, X. and Gelfand, A.~E. (2011).
\newblock The {D}irichlet labeling process for clustering functional data.
\newblock {\em Statistica Sinica}, pages 1249--1289.

\bibitem[Nieto-Barajas and Contreras-Crist{\'a}n, 2014]{nieto2014bayesian}
Nieto-Barajas, L.~E. and Contreras-Crist{\'a}n, A. (2014).
\newblock A {B}ayesian nonparametric approach for time series clustering.
\newblock {\em Bayesian Analysis}, 9(1):147--170.

\bibitem[Nobile, 2004]{nobile2004posterior}
Nobile, A. (2004).
\newblock On the posterior distribution of the number of components in a finite mixture.
\newblock {\em Annals of Statistics}, 32:2044--2073.

\bibitem[O’Hagan et~al., 2016]{o2016clustering}
O’Hagan, A., Murphy, T.~B., Gormley, I.~C., McNicholas, P.~D., and Karlis, D. (2016).
\newblock Clustering with the multivariate normal inverse {G}aussian distribution.
\newblock {\em Computational Statistics \& Data Analysis}, 93:18--30.

\bibitem[Paci et~al., 2013]{paci2013spatio}
Paci, L., Gelfand, A.~E., and Holland, D.~M. (2013).
\newblock Spatio-temporal modeling for real-time ozone forecasting.
\newblock {\em Spatial Statistics}, 4:79--93.

\bibitem[Page and Quintana, 2016]{page2016spatial}
Page, G.~L. and Quintana, F.~A. (2016).
\newblock Spatial {P}roduct {P}artition {M}odels.
\newblock {\em Bayesian Analysis}, 11(1):265--298.

\bibitem[Page et~al., 2022]{page2022dependent}
Page, G.~L., Quintana, F.~A., and Dahl, D.~B. (2022).
\newblock Dependent modeling of temporal sequences of random partitions.
\newblock {\em Journal of Computational and Graphical Statistics}, 31(2):614--627.

\bibitem[Palla et~al., 2005]{palla2005uncovering}
Palla, G., Der{\'e}nyi, I., Farkas, I., and Vicsek, T. (2005).
\newblock Uncovering the overlapping community structure of complex networks in nature and society.
\newblock {\em Nature}, 435(7043):814--818.

\bibitem[Papaspiliopoulos and Roberts, 2008]{papaspiliopoulos2008retrospective}
Papaspiliopoulos, O. and Roberts, G.~O. (2008).
\newblock {Retrospective Markov chain Monte Carlo methods for Dirichlet process hierarchical models}.
\newblock {\em Biometrika}, 95(1):169--186.

\bibitem[Paton and McNicholas, 2020]{paton2020detecting}
Paton, F. and McNicholas, P.~D. (2020).
\newblock Detecting {British Columbia} coastal rainfall patterns by clustering {G}aussian processes.
\newblock {\em Environmetrics}, 31(8):e2631.

\bibitem[Peluso et~al., 2020]{peluso2020bayesian}
Peluso, S., Mira, A., Rue, H., Tierney, N.~J., Benvenuti, C., Cianella, R., Caputo, M.~L., and Auricchio, A. (2020).
\newblock A {B}ayesian spatiotemporal statistical analysis of out-of-hospital cardiac arrests.
\newblock {\em Biometrical Journal}, 62(4):1105--1119.

\bibitem[Petralia et~al., 2012]{petralia2012repulsive}
Petralia, F., Rao, V., and Dunson, D. (2012).
\newblock Repulsive mixtures.
\newblock {\em Advances in neural information processing systems}, 25.

\bibitem[Pietrogrande et~al., 2022]{pietrogrande2022seasonal}
Pietrogrande, M.~C., Demaria, G., Colombi, C., Cuccia, E., and Dal~Santo, U. (2022).
\newblock Seasonal and spatial variations of {PM}10 and {PM}2.5 oxidative potential in five urban and rural sites across {L}ombardia {R}egion, {I}taly.
\newblock {\em International Journal of Environmental Research and Public Health}, 19(13):7778.

\bibitem[Pitman, 1995]{pitman1995exchangeable}
Pitman, J. (1995).
\newblock Exchangeable and partially exchangeable random partitions.
\newblock {\em Probability theory and related fields}, 102(2):145--158.

\bibitem[Pitman, 1996]{pitman1996}
Pitman, J. (1996).
\newblock Some developments of the {Blackwell-Macqueen URN} scheme.
\newblock {\em Lecture Notes-Monograph Series}, 30:245--267.

\bibitem[Quintana, 2006]{quintana2006predictive}
Quintana, F.~A. (2006).
\newblock A predictive view of {B}ayesian clustering.
\newblock {\em Journal of Statistical Planning and Inference}, 136(8):2407--2429.

\bibitem[Ranalli and Maruotti, 2020]{ranalli2020model}
Ranalli, M. and Maruotti, A. (2020).
\newblock Model-based clustering for noisy longitudinal circular data, with application to animal movement.
\newblock {\em Environmetrics}, 31(2):e2572.

\bibitem[Rasmussen and Williams, 2005]{williams2006gaussian}
Rasmussen, C.~E. and Williams, C. K.~I. (2005).
\newblock {\em {Gaussian Processes for Machine Learning}}.
\newblock The MIT Press.

\bibitem[Reich and Fuentes, 2007]{reich2007multivariate}
Reich, B.~J. and Fuentes, M. (2007).
\newblock A multivariate semiparametric {B}ayesian spatial modeling framework for hurricane surface wind fields.
\newblock {\em The Annals of Applied Statistics}, 1(1):249--264.

\bibitem[Richardson and Green, 1997]{richardson1997bayesian}
Richardson, S. and Green, P.~J. (1997).
\newblock On {B}ayesian analysis of mixtures with an unknown number of components (with discussion).
\newblock {\em Journal of the Royal Statistical Society Series B: Statistical Methodology}, 59(4):731--792.

\bibitem[Rodr{\i}guez and Dunson, 2011]{rodriguez2011nonparametric}
Rodr{\i}guez, A. and Dunson, D.~B. (2011).
\newblock Nonparametric {B}ayesian models through probit stick-breaking processes.
\newblock {\em Bayesian Analysis}, 6(1):145--178.

\bibitem[Rousseau et~al., 2019]{rousseau2019bayesian}
Rousseau, J., Grazian, C., and Lee, J.~E. (2019).
\newblock Bayesian mixture models: Theory and methods.
\newblock In {\em Handbook of mixture analysis}, pages 53--72. Chapman and Hall/CRC.

\bibitem[Rousseau and Mengersen, 2011]{rousseau2011asymptotic}
Rousseau, J. and Mengersen, K. (2011).
\newblock Asymptotic behaviour of the posterior distribution in overfitted mixture models.
\newblock {\em Journal of the Royal Statistical Society Series B: Statistical Methodology}, 73(5):689--710.

\bibitem[Rushworth et~al., 2017]{rushworth2017adaptive}
Rushworth, A., Lee, D., and Sarran, C. (2017).
\newblock An adaptive spatiotemporal smoothing model for estimating trends and step changes in disease risk.
\newblock {\em Journal of the Royal Statistical Society Series C: Applied Statistics}, 66(1):141--157.

\bibitem[Sahu et~al., 2006]{sahu2006}
Sahu, S.~K., Gelfand, A.~E., and M, D. (2006).
\newblock Spatio-temporal modeling of fine particulate matter.
\newblock {\em Journal of Agricultural, Biological, and Environmental Statistics}, 11:61--86.

\bibitem[Sarang, 2023]{sarang2023centroid}
Sarang, P. (2023).
\newblock {\em Centroid-Based Clustering}, pages 171--183.
\newblock Springer International Publishing, Cham.

\bibitem[Sethuraman, 1994]{sethuraman1994constructive}
Sethuraman, J. (1994).
\newblock A constructive definition of {D}irichlet priors.
\newblock {\em Statistica sinica}, pages 639--650.

\bibitem[Stephens, 2000a]{stephens2000bayesian}
Stephens, M. (2000a).
\newblock Bayesian analysis of mixture models with an unknown number of components--an alternative to reversible jump methods.
\newblock {\em The Annals of Statistics}, 28(1):40--74.

\bibitem[Stephens, 2000b]{stephens2000dealing}
Stephens, M. (2000b).
\newblock Dealing with label switching in mixture models.
\newblock {\em Journal of the Royal Statistical Society: Series B (Statistical Methodology)}, 62(4):795--809.

\bibitem[Stroud et~al., 2001]{stroud2001dynamic}
Stroud, J.~R., M{\"u}ller, P., and Sans{\'o}, B. (2001).
\newblock Dynamic models for spatiotemporal data.
\newblock {\em Journal of the Royal Statistical Society: Series B (Statistical Methodology)}, 63(4):673--689.

\bibitem[Torabi, 2014]{torabi2014spatiotemporal}
Torabi, M. (2014).
\newblock Spatiotemporal modeling of odds of disease.
\newblock {\em Environmetrics}, 25(5):341--350.

\bibitem[Vanhatalo et~al., 2021]{vanhatalo2021spatiotemporal}
Vanhatalo, J., Foster, S.~D., and Hosack, G.~R. (2021).
\newblock Spatiotemporal clustering using {G}aussian processes embedded in a mixture model.
\newblock {\em Environmetrics}, 32(7):e2681.

\bibitem[Wade, 2023]{wade2023bayesian}
Wade, S. (2023).
\newblock Bayesian cluster analysis.
\newblock {\em Philosophical Transactions of the Royal Society A}, 381(2247):20220149.

\bibitem[Wade and Ghahramani, 2018]{wade2018bayesian}
Wade, S. and Ghahramani, Z. (2018).
\newblock Bayesian cluster analysis: Point estimation and credible balls (with discussion).
\newblock {\em Bayesian Analysis}, 13(2):559--626.

\bibitem[Waller et~al., 1997]{waller1997hierarchical}
Waller, L.~A., Carlin, B.~P., Xia, H., and Gelfand, A.~E. (1997).
\newblock Hierarchical spatio-temporal mapping of disease rates.
\newblock {\em Journal of the American Statistical association}, 92(438):607--617.

\bibitem[Wan et~al., 2021]{wan2021spatio}
Wan, Y., Xu, M., Huang, H., and Xi~Chen, S. (2021).
\newblock A spatio-temporal model for the analysis and prediction of fine particulate matter concentration in {B}eijing.
\newblock {\em Environmetrics}, 32(1):e2648.

\bibitem[Wang et~al., 2024]{wang2024spatiotemporal}
Wang, F., Duan, C., Li, Y., Huang, H., and Shia, B.-C. (2024).
\newblock Spatiotemporal varying coefficient model for respiratory disease mapping in {T}aiwan.
\newblock {\em Biostatistics}, 25(1):40--56.

\bibitem[West and Harrison, 2006]{west2006bayesian}
West, M. and Harrison, J. (2006).
\newblock {\em Bayesian forecasting and dynamic models}.
\newblock Springer Science \& Business Media.

\bibitem[{World Health Organization}, 2022]{who2022world}
{World Health Organization} (2022).
\newblock {\em World health statistics 2022: monitoring health for the SDGs, sustainable development goals}.
\newblock World Health Organization.

\bibitem[Wu and Luo, 2022]{wu2022nonparametric}
Wu, Q. and Luo, X. (2022).
\newblock Nonparametric {B}ayesian two-level clustering for subject-level single-cell expression data.
\newblock {\em Statistica Sinica}, 32(4):1835--1856.

\bibitem[Xie and Xu, 2020]{xie2020bayesian}
Xie, F. and Xu, Y. (2020).
\newblock Bayesian repulsive {G}aussian mixture model.
\newblock {\em Journal of the American Statistical Association}, 115(529):187--203.

\bibitem[Xu et~al., 2016]{xu2016bayesian}
Xu, Y., M{\"u}ller, P., and Telesca, D. (2016).
\newblock Bayesian inference for latent biologic structure with determinantal point processes ({DPP}).
\newblock {\em Biometrics}, 72(3):955--964.

\end{thebibliography}

\appendix

\begin{appendices}
\renewcommand{\thesection}{A\arabic{section}}
\renewcommand{\theequation}{A\arabic{equation}}
\renewcommand{\thefigure}{A\arabic{figure}}

\setcounter{equation}{0}
\setcounter{figure}{0}

\section*{Appendices}
\addcontentsline{toc}{section}{Appendices} 

\section{Connections between mixture models and clustering}\label{app:clust_mix}

We aim to illustrate how the mixture model is intrinsically linked to a clustering model. First, we recall that  $P \equiv (\boldsymbol{\gamma}, M, \bs{\pi})$, then we consider the  formulation of the mixture model given in Equation \eqref{eq:mod_clust}. So, the joint model for the data and parameters is given by 
\begin{equation*}
    \mathcal{L}(\bs{y},\bs{c},\bs{\gamma},\bs{\pi}) = \prod_{i=1}^n f(y_i \mid \gamma_{c_i}) \prod_{i=1}^n \pi_{c_i} p(\bs{\pi}) \prod_{m\geq 1} p_0(\gamma_m), 
\end{equation*}
where $p_0(\gamma_m)$ is the density or the probability mass function of the parameters, and the prior for $M$ is included in $p(\bs{\pi})$.

First we let $\theta_i=\gamma_{c_i}$ for $i=1,\dots,n$ so that $\theta_i|P\stackrel{iid}{\sim}P$. Since $P$ is almost sure discrete, there will be ties among the $\theta_i$'s.  We define $\theta_k^\ast$ for $k=1,\dots,K$ as the unique values in $\theta_1,\dots,\theta_n$. This observation is crucial for clustering, which now can be represented via the partitions $\rho_n = \{S_1,\dots,S_K\}$, where observation $i \in S_k$ if and only if $\theta_i = \theta^\ast_k$, with $n_k = \left|S_k\right|$. Let's now denote by $c^\ast_k$ the values among $\{1,2,\dots, M\}$ such that $\theta_k^\ast=\gamma_{c^\ast_k}$. So that, given $\boldsymbol{\theta}^\ast=(\theta_1^\ast,\dots,\theta_K^\ast)$, we obtain  
\begin{equation*}
    \mathcal{L}(\bs{y},\bs{\theta}^\ast,\rho_n,\bs{\gamma},\bs{\pi},c^\ast_1,\dots,c^\ast_K) = \prod_{k=1}^K \prod_{i \in S_k} f(y_i \mid \theta_k^\ast) \pi_{c_k^\ast}^{n_k} p(\bs{\pi}) \prod_{m\geq 1} p_0(\gamma_m)\delta_{\theta_k^\ast}(\gamma_{c^\ast_k}). 
\end{equation*}  
Marginalizing over $\gamma_m$ for all $m\ge 1$, we obtain 
\begin{equation*}
    \mathcal{L}(\bs{y},\bs{\theta}^\ast,\rho_n,\bs{\pi},c^\ast_1,\dots,c^\ast_K) = \prod_{k=1}^K \prod_{i \in S_k} f(y_i \mid \gamma_{c_k^\ast}) \, \pi_{c_k^\ast}^{n_k}\, p_0(\theta^\ast_k) \, p(\bs{\pi}) .
\end{equation*}  

Next, we marginalize over the specific values of $c_1^\ast,\dots,c_K^\ast$ and weights $\bs{\pi}$. Then we can conclude that:  
\begin{align}\label{eq:eppf_aux}
        \mathcal{L}(\bs{y}, \rho_n,\theta_1^\ast,\dots,\theta_K^\ast) & =\prod_{k=1}^K\prod_{i \in S_k} f(y_i \mid \theta_{k}^\ast) \, p_0(\theta^\ast_k)
        \int_{\mathcal{S}_M} 
        \sum_{c_1^\ast,\dots, c_K^\ast} \pi_{c_k^\ast}^{n_k} p(\bs{\pi}) \notag \\
        &= \prod_{k=1}^K\prod_{i \in S_k} f(y_i \mid \theta_{k}^\ast)\, p_0(\theta^\ast_k) \, \mathbb{E}\left[ \sum_{c_1^\ast,\dots, c_K^\ast} \pi_{c_k^\ast}^{n_k} \right],
\end{align}  
Where $\mathcal{S}_M$ is the $(M-1)$-dimensional simplex, i.e,. the space over which the vector of probability weights $\bs{\pi}$ lives, while the summation in the expectation is taken over all sequences of positive integers, i.e., $c_1^\ast, \dots,c_K^\ast \in \mathbb{N}$ with $c_k^\ast \neq c_{k^\prime}^\ast$ for $k \neq k^\prime$. The term $\mathbb{E}\left( \sum_{c_1^\ast,\dots, c_K^\ast} \pi_{c_k^\ast}^{n_k} \right)$ is known as the exchangeable partition probability function (EPPF), it is the prior over the partition induced by the process $P$. So we can rewrite expression \eqref{eq:eppf_aux}
as 
\begin{align}\label{eq:eppf_aux2}
        \mathcal{L}(\bs{y}, \rho_n,\theta_1^\ast,\dots,\theta_K^\ast) &= \prod_{k=1}^K\prod_{i \in S_k} f(y_i \mid \theta_{k}^\ast)\, p_0(\theta^\ast_k) \, \text{EPPF}(n_1,\dots,n_k).
\end{align} 
Now it is trivial to observe that Expression \ref{eq:eppf_aux2} is the joint law of data and parameter given by model  \eqref{eq:mod_clust_eppf}. And this proves that Model \eqref{eq:mod_clust_eppf} is obtained by marginalizing $P$ from model \eqref{eq:mod_clust}. To conclude, we mention that \cite{pitman1995exchangeable} demonstrated that the converse also holds. Namely, that if we assign a clustering model as in \eqref{eq:mod_clust_eppf} by assuming a proper EPPF, then there exist an almost sure random probability measure (the EPPF identifies which law) such that the model can be expressed as a mixture in Equation \eqref{eq:mod_clust}.

\section{Similarity function}
\label{app:similarity}
Here we provide the computations through which we obtained the closed form of the similarity function $g_3(\cdot)$ in \eqref{eq:similarity3}:
{\allowdisplaybreaks
\begin{align*}
    & \prod_{i \in S_{k}} q\left(\bs{s}_{i} \mid \bs{\xi}_{k}\right) q\left(\bs{\xi}_{k}\right) =  \prod_{i \in S_{k}} \mathcal{N}_{2}\left(\bs{s}_{i} \mid \bs{m}_{k}, V_{k}\right) N I W\left(\bs{m}_{k}, V_{k} \mid \bs{\mu}_{0}, \kappa_{0}, \nu_{0}, \Lambda_{0}\right) \\
    = & \prod_{i \in S_{k}}(2 \pi)^{-\frac{D}{2}}\left|V_{k}\right|^{-1 /2} \exp \left(-\frac{1}{2}\left(\bs{s}_{i}-\bs{m}_{k}\right)^{\top} V_{k}^{-1}\left(\bs{s}_{i}-\bs{m}_{k}\right)\right) \\
    &\cdot \frac{\kappa_{0}^{\frac{D}{2}}\left|\Lambda_{0}\right|^{\frac{\nu_{0}}{2}}\left|V_{k}\right|^{-\frac{\nu_{0}+D+2}{2}}}{(2 \pi)^{\frac{D}{2}} 2^{\frac{\nu_{0}D}{2}} \Gamma_{D}\left(\frac{\nu_{0}}{2}\right)} \exp \left(-\frac{\kappa_{0}}{2}\left(\bs{m}_{k}-\bs{\mu}_{0}\right)^{\top} V_{k}^{-1}\left(\bs{m}_{k}-\bs{\mu}_{0}\right)-\frac{1}{2} \operatorname{tr}\left(\Lambda_{0} V_{k}^{-1}\right)\right) \\
    = & \frac{\kappa_{0}^{\frac{D}{2}}\left|\Lambda_{0}\right|^{\frac{\nu_{0}}{2}}\left|V_{k}\right|^{-\frac{(\nu_{0}+n_{k})+D+2}{2}}}{(2 \pi)^{(\left| S_{k} \right| + 1)\frac{D}{2}} 2^{\frac{\nu_{0}D}{2}} \Gamma_{D}\left(\frac{\nu_{0}}{2}\right)} 
    \begin{aligned}[t]
        \exp \biggl(&-\frac{1}{2} \sum_{i \in S_{k}}\left(\bs{s}_{i}-\bs{m}_{k}\right)^{\top} V_{k}^{-1}\left(\bs{s}_{i}-\bs{m}_{k}\right)\\
        &-\frac{\kappa_{0}}{2}\left(\bs{m}_{k}-\bs{\mu}_{0}\right)^{\top} V_{k}^{-1}\left (\bs{m}_{k}-\bs{\mu}_{0}\right) - \frac{1}{2} \operatorname{tr}\left(\Lambda_{0} V_{k}^{-1}\right)\biggr)
    \end{aligned}\\
    = &\frac{\kappa_{0}^{\frac{D}{2}}\left|\Lambda_{0}\right|^{\frac{\nu_{0}}{2}}\left|V_{k}\right|^{-\frac{(\nu_{0}+n_{k})+D+2}{2}}}{(2 \pi)^{(\left| S_{k} \right| + 1)\frac{D}{2}} 2^{\frac{\nu_{0}D}{2}} \Gamma_{D}\left(\frac{\nu_{0}}{2}\right)} 
    \begin{aligned}[t]
        \exp \biggl(&-\frac{n_{k}}{2}\left(\bs{m}_{k}-\Bar{\bs{s}}_{k}\right)^{\top} V_{k}^{-1}\left(\bs{m}_{k}-\Bar{\bs{s}}_{k}\right) \\ 
        &-\frac{\kappa_{0}}{2}\left(\bs{m}_{k}-\bs{\mu}_{0}\right)^{\top} V_{k}^{-1}\left (\bs{m}_{k}-\bs{\mu}_{0}\right) \\
        & -\frac{1}{2} \operatorname{tr}\left(\Lambda_{0} V_{k}^{-1}\right) -\frac{1}{2} \operatorname{tr}\left(\Delta_{k} V_{k}^{-1}\right) \biggr)
    \end{aligned}\\
    = &\frac{\kappa_{0}^{\frac{D}{2}}\left|\Lambda_{0}\right|^{\frac{\nu_{0}}{2}}\left|V_{k}\right|^{-\frac{(\nu_{0}+n_{k})+D+2}{2}}}{(2 \pi)^{(\left| S_{k} \right| + 1)\frac{D}{2}} 2^{\frac{\nu_{0}D}{2}} \Gamma_{D}\left(\frac{\nu_{0}}{2}\right)} 
    \begin{aligned}[t]
        \exp \biggl(&-\frac{1}{2}\left(\bs{m}_{k}-\frac{n_{k}\Bar{\bs{s}}_{k} + \kappa_{0} \bs{\mu}_{0}}{n_{k} + \kappa_{0}}\right)^{\top}\left(n_{k} + \kappa_{0}\right) V_{k}^{-1}\left(\bs{m}_{k}-\frac{n_{k}\Bar{\bs{s}}_{k} + \kappa_{0} \bs{\mu}_{0}}{n_{k} + \kappa_{0}}\right) \\
        & - \frac{1}{2} \left(\bs{\mu}_{0}-\Bar{\bs{s}}_{k}\right)^{\top} \frac{\kappa_{0}n_{k}}{n_{k}+\kappa_{0}} V_{k}^{-1} \left(\bs{\mu}_{0}-\Bar{\bs{s}}_{k}\right) \\
        & - \frac{1}{2} \operatorname{tr}\left(\Lambda_{0} V_{k}^{-1}\right) -\frac{1}{2} \operatorname{tr}\left(\Delta_{k} V_{k}^{-1}\right) \biggr)
    \end{aligned}\\
    = &\frac{\kappa_{0}^{\frac{D}{2}}\left|\Lambda_{0}\right|^{\frac{\nu_{0}}{2}}\left|V_{k}\right|^{-\frac{(\nu_{0}+n_{k})+D+2}{2}}}{(2 \pi)^{(\left| S_{k} \right| + 1)\frac{D}{2}} 2^{\frac{\nu_{0}D}{2}} \Gamma_{D}\left(\frac{\nu_{0}}{2}\right)} 
    \begin{aligned}[t]
        \exp \biggl(&-\frac{1}{2}\left(\bs{m}_{k}-\frac{n_{k}\Bar{\bs{s}}_{k} + \kappa_{0} \bs{\mu}_{0}}{n_{k} + \kappa_{0}}\right)^{\top}\left(n_{k} + \kappa_{0}\right) V_{k}^{-1}\left(\bs{m}_{k}-\frac{n_{k}\Bar{\bs{s}}_{k} + \kappa_{0} \bs{\mu}_{0}}{n_{k} + \kappa_{0}}\right) \\
        & -\frac{1}{2} \operatorname{tr}\left(\left(\Lambda_{0}+\Delta_{k}+\frac{\kappa_{0}n_{k}}{n_{k}+\kappa_{0}}\left(\Bar{\bs{s}}_{k}-\bs{\mu}_{0} \right)\left(\Bar{\bs{s}}_{k}-\bs{\mu}_{0}\right)^{\top}\right)V_{k}^{-1}\right)\biggr).
    \end{aligned}
\end{align*}
}
Simplifying the notation:
\begin{align*}
    \prod_{i \in S_{k}} q\left(\bs{s}_{i} \mid \bs{\xi}_{k}\right) q\left(\bs{\xi}_{k}\right) = 
    \frac{\kappa_{0}^{\frac{D}{2}}\left|\Lambda_{0}\right|^{\frac{\nu_{0}}{2}}\left|V_{k}\right|^{-\frac{\nu_{k}+D+2}{2}}}{(2 \pi)^{(\left| S_{k} \right| + 1)\frac{D}{2}} 2^{\frac{\nu_{0}D}{2}} \Gamma_{D}\left(\frac{\nu_{0}}{2}\right)} \exp \biggl(&-\frac{\kappa_{k}}{2}\left(\bs{m}_{k}-\bs{\mu}_{k}\right)^{\top}V_{k}^{-1}\left(\bs{m}_{k}-\bs{\mu}_{k}\right) \\ 
    &- \frac{1}{2} \operatorname{tr}\left(\Lambda_{k}V_{k}^{-1}\right)\biggr),
\end{align*}
where $\bs{\mu}_{k} =  \frac{n_{k}\Bar{\bs{s}}_{k} + \kappa_{0} \bs{\mu}_{0}}{n_{k} + \kappa_{0}}$. 

The similarity function then becomes:
\begin{align*}
    C\left(S_{k}^{(-i)} \cup \{i\}\right)g\left(s_{k}^{{\ast}(-i)} \cup \bs{s}_{i} \right) =& \alpha \, \Gamma\left(n_{k}\right) \, \int \prod_{i \in S_{k}} q\left(\bs{s}_{i} \mid \bs{\xi}_{k}\right) q\left(\bs{\xi}_{k}\right) d \bs{\xi}_{k} \\
    =& \alpha \, \Gamma\left(n_{k}\right) \, \frac{1}{(2\pi)^{\left| S_{k} \right|\frac{D}{2}}} \frac{\kappa_{0}^{\frac{D}{2}}\left|\Lambda_{0}\right|^{\frac{\nu_{0}}{2}} 2^{\frac{\nu_{k}D}{2}} \Gamma_{D}\left(\frac{\nu_{k}}{2}\right)}{ \kappa_{k}^{\frac{D}{2}} \left|\Lambda_{k}\right|^{\frac{\nu_{k}}{2}} 2^{\frac{\nu_{0}D}{2}} \Gamma_{D}\left(\frac{\nu_{0}}{2}\right)},
\end{align*}
so that
\begin{align*}
    \frac{C\left(S_{k}^{(-i)} \cup \{i\}\right)g\left(s_{k}^{{\ast}(-i)} \cup \bs{s}_{i} \right)}{C\left(S_{k}^{(-i)}\right)g\left(s_{k}^{{\ast}(-i)}\right)} = \frac{\Gamma\left(n_{k}\right)}{\Gamma\left(\left|S_{k}^{(-i)}\right|\right)} \cdot \frac{(2\pi)^{\left| S_{k}^{(-i)} \right|\frac{D}{2}}{\kappa_{k}^{(-i)}}^{\frac{D}{2}} \left|\Lambda_{k}^{(-i)}\right|^{\frac{\nu_{k}^{(-i)}}{2}} 2^{\frac{\nu_{k}D}{2}} \Gamma_{D}\left(\frac{\nu_{k}}{2}\right)}{(2\pi)^{\left| S_{k} \right|\frac{D}{2}} \kappa_{k}^{\frac{D}{2}} \left|\Lambda_{k}\right|^{\frac{\nu_{k}}{2}}2^{\frac{\nu_{k}^{(-i)}D}{2}} \Gamma_{D}\left(\frac{\nu_{k}^{(-i)}}{2}\right)}, 
\end{align*}
while 
\begin{align*}
    C\left(\{i\}\right)g\left(\bs{s}_{i} \right) = \frac{\alpha \, \Gamma\left(\left|\{i\}\right|\right) }{(2\pi)^\frac{D}{2}} \cdot\frac{\kappa_{0}^{\frac{D}{2}}\left|\Lambda_{0}\right|^{\frac{\nu_{0}}{2}} 2^{\frac{(\nu_{0}+1)D}{2}} \Gamma_{D}\left(\frac{\nu_{0}+1}{2}\right)}{(\kappa_{0} + 1)^{\frac{D}{2}} \left|\Lambda_{0} + \frac{\kappa_{0}}{1+\kappa_{0}}\left(\bs{s}_{i}-\bs{\mu}_{0} \right)\left(\bs{s}_{i}-\bs{\mu}_{0}\right)^{\top} \right|^{\frac{\nu_{0}+1}{2}} 2^{\frac{\nu_{0}D}{2}} \Gamma_{D}\left(\frac{\nu_{0}}{2}\right)},
\end{align*}
where $\Bar{\bs{s}}_{k}^{(-i)}$ is the mean of the elements belonging to $S_{k}^{(-i)}$ and 
{\allowdisplaybreaks
\begin{align*}
    \Delta_{k}^{(-i)} &= \sum_{j \in S_{k}^{(-i)}}\left(\bs{s}_{j}-\Bar{\bs{s}}_{k}^{(-i)}\right)\left(\bs{s}_{j}-\Bar{\bs{s}}_{k}^{(-i)}\right)^{\top}, \\
    \kappa_{k}^{(-i)} &= \kappa_{0} + \left|S_{k}^{(-i)}\right|, \\
    \nu_{k}^{(-i)} &= \nu_{0} + \left|S_{k}^{(-i)}\right|, \\
    \Lambda_{k}^{(-i)} &= \left(\Lambda_{0}+\Delta_{k}^{(-i)}+\frac{\kappa_{0}\left|S_{k}^{(-i)}\right|}{\left|S_{k}^{(-i)}\right|+\kappa_{0}}\left(\Bar{\bs{s}}_{k}^{(-i)}-\bs{\mu}_{0} \right)\left(\Bar{\bs{s}}_{k}^{(-i)}-\bs{\mu}_{0}\right)^{\top}\right).
\end{align*}}

\section{Computational details}\label{app:comp_det}

Posterior sampling, described in Algorithm~\ref{alg:sppm_gs}, requires evaluating the density of $\bs{w}_i$ at the cluster-specific parameters. This step represents a significant computational bottleneck, as it involves calculating a quadratic form and a log-determinant, both involving the $T \times T$ precision matrix $\bs{R}(\bs{\theta}_{i})^{-1}$. For a generic matrix of size $T \times T$, such operations have computational costs of $\mathcal{O}(T^2)$ and $\mathcal{O}(T^3)$, respectively. However, we achieve a more efficient implementation by leveraging the tridiagonal structure of $\bs{R}(\bs{\theta}_{i})^{-1}$, namely
\begin{equation*}
    \bs{R}(\bs{\theta}_{i})^{-1} = \frac{1}{\tau_{i}^2}
    \begin{bmatrix}
        1 & -\phi_i & 0 & \cdots & 0 \\
        -\phi_i & 1 + \phi_i^2 & -\phi_i & \cdots & 0 \\
        0 & -\phi_i & 1 + \phi_i^2 & \ddots & \vdots \\
        \vdots & \vdots & \ddots & \ddots & -\phi_i \\
        0 & 0 & \cdots & -\phi_i & 1 
    \end{bmatrix}.
\end{equation*}
As a consequence, the quadratic form associated with $\bs{R}(\bs{\theta}_{i})^{-1}$ is given by
\begin{align} \label{eq:fast_quadratic}
    \bs{w}_{i}^\top \bs{R}(\bs{\theta}_{i})^{-1} \bs{w}_{i} =  \frac{w_{i1}^2 + w_{iT}^2}{\tau_{i}^2} + \frac{1 + \phi_{i}^2}{\tau_{i}^2} \sum_{t=2}^{T-1} w_{it}^2 - \frac{2\phi_i}{\tau_{i}^2} \sum_{t=1}^{T-1} w_{it}w_{i(t+1)}.
\end{align}
This approach allows to eliminate unnecessary computations involving the zeros in $\bs{R}(\bs{\theta}_{i})^{-1}$, thus leading to a reduced computational cost of $\mathcal{O}(T)$, instead of $\mathcal{O}(T^2)$, for the quadratic form in \eqref{eq:fast_quadratic}. 

For computing the log-determinant of $\bs{R}(\bs{\theta}_{i})^{-1}$ while avoiding numerical issues, we employed a two-fold strategy following \cite{meurant1992review}. First, we calculate the Cholesky decomposition of $\bs{R}(\bs{\theta}_{i})^{-1} = \bs{L}_{i} \bs{L}_{i}^\top$, which can be obtained in closed form. We then take advantage of the lower triangular structure of $\bs{L}_{i}$ to efficiently compute the determinant. More precisely,
\begin{equation*}
    \bs{L}_{i} =
    \begin{bmatrix}
        \sqrt{\delta_1} & 0 & 0 & \cdots & 0 \\
        -\phi_i\tau_{i}^{-2} \delta_1^{-1/2} & \sqrt{\delta_2} & 0 & \cdots & 0 \\ 
        0 & -\phi_i\tau_{i}^{-2} \delta_2^{-1/2} & \sqrt{\delta_3} & \ddots & \vdots \\
        \vdots & \vdots & \ddots & \ddots & 0 \\
        0 & 0 & \cdots & -\phi_i\tau_{i}^{-2} \delta_{T-1}^{-1/2} & \sqrt{\delta_T} 
    \end{bmatrix},
\end{equation*}
with 
\begin{equation*}
    \delta_t =
    \begin{cases}
        \tau_i^{-2} &\quad t = 1, \\
        (1+\phi_i^2)\tau_i^{-2} - \phi_i^2\tau_i^{-4}\delta_{t-1}^{-1} &\quad t \in \{2,\dots,T-1\}, \\
        \tau_i^{-2} - \phi_i^2\tau_i^{-4}\delta_{T-1}^{-1} &\quad t = T.
    \end{cases}
\end{equation*}
Hence,
\begin{equation} \label{eq:fast_logdet}
    \log \left| \bs{R}(\bs{\theta}_{i})^{-1} \right| = \log \left| \bs{L}_{i} \right|^2 = 2 \log \left(\prod_{t=1}^T \sqrt{\delta_t}\right) = 2\sum_{t=1}^T \log \sqrt{\delta_t} = \sum_{t=1}^T \log \delta_t.
\end{equation}
The formula in~\eqref{eq:fast_logdet} allows to reduce the computational cost of computing the log-determinant of the precision matrix from $\mathcal{O}(T^3)$ to $\mathcal{O}(T)$.

Another computational issue regards sampling from the full conditional distribution of $\bs{w}_{i}$, that is
\begin{equation}\label{eq:full_cond_w}
    \bs{w}_{i} \mid \sigma_{i}^{2}, \bs{\theta}_{i}, \bs{\beta}_{i}, \bs{Z}, \bs{y}_{i} \sim \mathcal{N}_T\left(\bs{V}_{\bs{w}_i}\bs{\mu}_{i}, \bs{V}_{\bs{w}_i} \right),
\end{equation}
where $\bs{\mu}_i = \sigma_{i}^{-2}(\bs{y}_{i} - \bs{Z}\bs{\beta}_i)$ and $\bs{V}_{\bs{w}_i}^{-1} = \sigma_{i}^{-2} \bs{I}_T + \bs{R}(\bs{\theta}_{i})^{-1}$. Namely, the inversion of a generic $T \times T$ precision matrix $\bs{V}_{\bs{w}_i}^{-1}$ has a computational cost of $\mathcal{O}(T^3)$. However, since $\bs{R}(\bs{\theta}_{i})^{-1}$ has a tridiagonal structure, $\bs{V}_{\bs{w}_i}^{-1}$ shares the same property. Following~\cite{mccausland2011simulation}, we exploit the tridiagonal structure of $\bs{V}_{\bs{w}_i}^{-1}$ for sampling efficiently from~\eqref{eq:full_cond_w}. To clarify, let $\bs{M}_{i}$ be a lower-triangular matrix derived from the Cholesky decomposition of the precision matrix $\bs{V}_{\bs{w}_i}^{-1} = \bs{M}_{i} \bs{M}_{i}^\top$. Consequently, $\bs{V}_{\bs{w}_i}=\bs{M}_{i}^{-\top} \bs{M}_{i}^{-1}$, where $\bs{M}_{i}^{-\top}$ denotes the inverse of the transpose of $\bs{M}_i$. Therefore, a random draw from \eqref{eq:full_cond_w} can be obtained through 
\begin{equation*}
    \bs{M}_{i}^{-\top} \bs{M}_{i}^{-1}\bs{\mu}_i + \bs{M}_{i}^{-\top} \bs{\kappa}_{i} = \bs{M}_{i}^{-\top} (\bs{M}_{i}^{-1}\bs{\mu}_i + \bs{\kappa}_{i}),
\end{equation*}
where $\bs{\kappa}_{i} \sim \mathcal{N}_T(0,\bs{I}_T)$. This leads to the following efficient algorithm for sampling from~\eqref{eq:full_cond_w}:
\begin{enumerate}
    \item Compute the Cholesky decomposition $\bs{V}_{\bs{w}_i}^{-1} = \bs{M}_{i} \bs{M}_{i}^\top$. 
    \item Sample $\bs{\kappa}_{i} \sim \mathcal{N}_T(0,\bs{I}_T)$. 
    \item Solve the linear equation $\bs{M}_{i} \bs{a} = \bs{\mu}_i$ for $\bs{a}$ using forward-substitution. 
    \item Solve the linear equation $\bs{M}_{i}^\top \bs{b} = \bs{a} + \bs{\kappa}_{i}$ for $\bs{b}$ using back-substitution. 
    \item Return $\bs{b}$ as a draw from \eqref{eq:full_cond_w}.
\end{enumerate}
The computational gains from this procedure are twofold. First, recognizing that the matrix $\bs{V}_{\bs{w}_i}^{-1}$ is tridiagonal allows for fast computation of the Cholesky decomposition via dedicated routines. Second, solving the linear equations at steps 3 and 4 of the algorithm above via forward and back substitution eliminates the need to compute the inverse of $\bs{M}_{i}$. Overall, this strategy reduces the computational cost of sampling from~\eqref{eq:full_cond_w} from $\mathcal{O}(T^3)$ to $\mathcal{O}(T)$, thus aligning with the recommendations of \cite{kastner2014ancillarity} and \cite{bitto2019achieving}.

\section{Additional results on the air quality  data analysis}\label{app:additional}
\begin{figure}[htbp!]
    \centering
    \includegraphics[width=\linewidth]{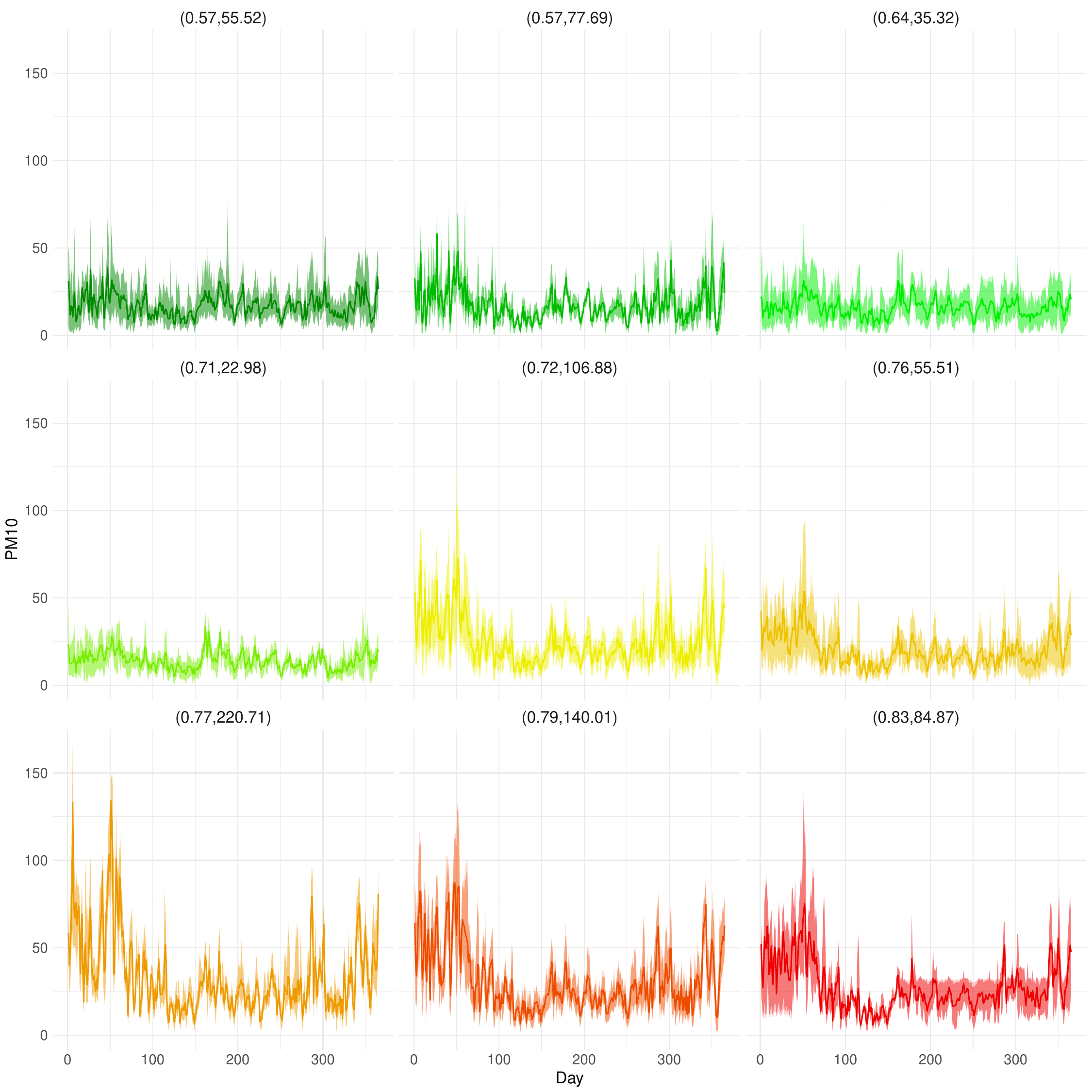}
    \caption{Mean (solid line), interquartile and 90\% interquantile bands for each of the obtained clusters resulting from PPM. On top of each panel, we report the estimated parameters $\bs{\theta}_{i} = (\phi_{i}, \tau_{i}^{2})$ within each cluster, under the PPM.}
    \label{fig:ts_comparison_cohesion}
\end{figure}

Each panel in Figure~\ref{fig:ts_comparison_cohesion} displays the time series of the average PM10 along with the 90\% interquantile bands, grouped according to the cluster membership estimated using the VI criterion under the PPM model. Each cluster comprises time series with similar temporal patterns, providing a representation of air pollution variability under various conditions. The panels illustrate the heterogeneity in PM10 trends across clusters. Some clusters exhibit relatively stable patterns with minor fluctuations, whereas others show pronounced variations, likely influenced by meteorological conditions, seasonal effects, or pollution sources. 

\begin{figure}[htbp!]
    \centering
    \includegraphics[width=\linewidth]{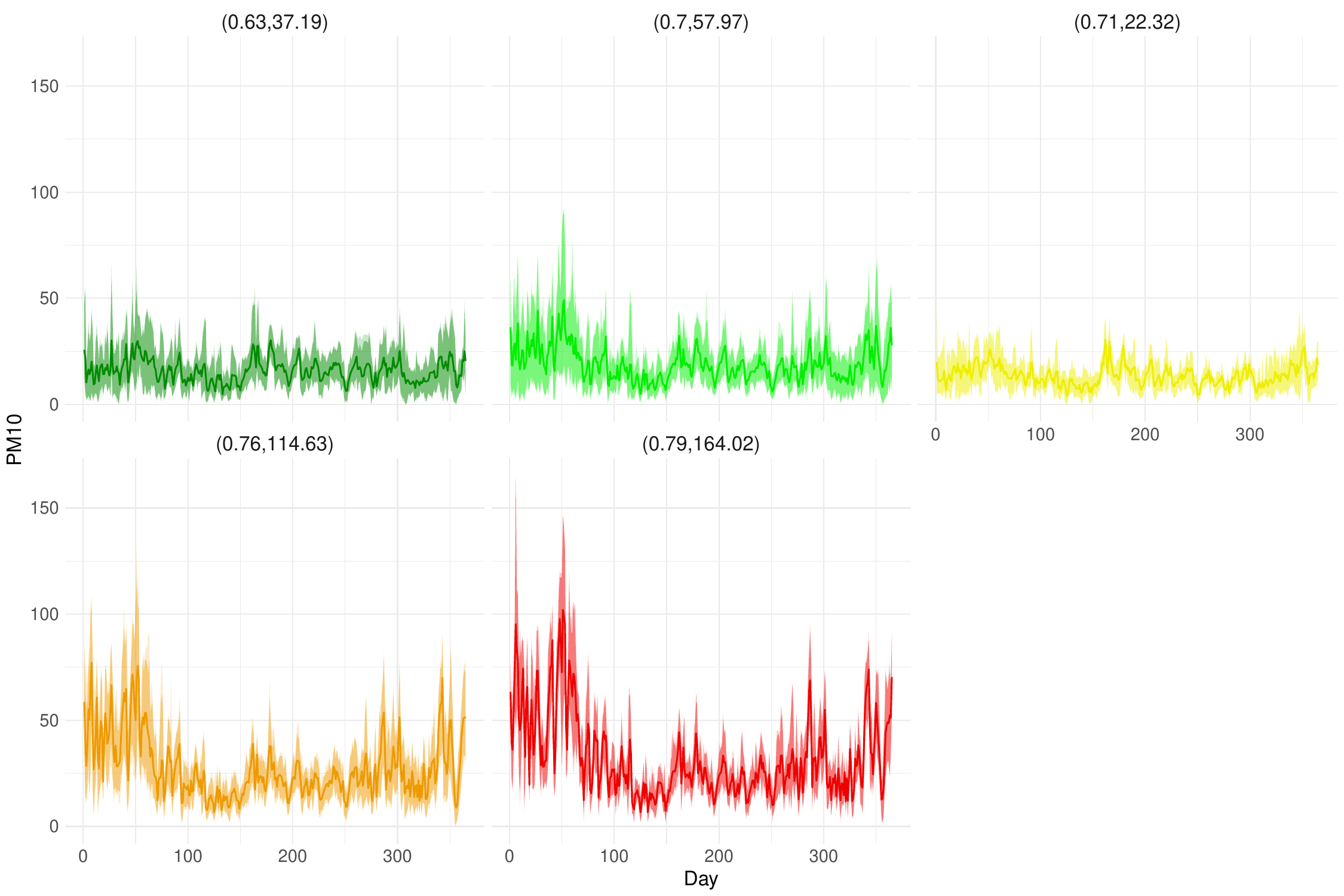}
    \caption{Mean (solid line), interquartile and 90\% interquantile bands for each of the obtained clusters resulting from sPPM. On top of each panel, we report the estimated parameters $\bs{\theta}_{i} = (\phi_{i}, \tau_{i}^{2})$ within each cluster, under the sPPM.}
    \label{fig:ts_comparison_similarity}
\end{figure}

Figure~\ref{fig:ts_comparison_similarity} presents the corresponding results obtained from the sPPM. As shown in the main manuscript, the sPPM yields a smaller number of clusters than the PPM, as it merges similar time series that are spatially close to each other. Despite the lower number of clusters, the main differences among the clusters remain consistent with Figure~\ref{fig:ts_comparison_cohesion}. This suggests that incorporating spatial similarity in clustering enhances interpretability by reducing redundancy without compromising the meaningful distinctions between clusters. As a result, the sPPM provides a more parsimonious yet equally informative representation of air pollution trends, making it a valuable tool for risk assessment and policy evaluation.

\end{appendices}

\end{document}